\documentclass[12pt, draftclsnofoot, onecolumn]{IEEEtran} 
\usepackage{cite} 

\usepackage[T1]{fontenc}    
\usepackage[utf8]{inputenc} 
\usepackage{newtxtext,newtxmath} 
\usepackage[english]{babel} 

\usepackage{amsmath, amsfonts, mathtools} 

\usepackage{graphicx}
\usepackage[caption=false, font=footnotesize]{subfig} 
\usepackage{booktabs}  
\usepackage{array}     
\usepackage{tabularx}  
\newcolumntype{Y}{>{\centering\arraybackslash}X} 

\usepackage{float}
\usepackage{lscape}
\usepackage{multirow}
\usepackage{adjustbox}
 \usepackage{tikz}
\usetikzlibrary{arrows.meta}
\usetikzlibrary{positioning}
\usetikzlibrary{calc}
\newcommand{\Mbig}{\mathsf{M}_{\mathrm{big}}}

\usepackage[ruled,vlined]{algorithm2e}
\RestyleAlgo{ruled}

\SetAlFnt{\footnotesize\rmfamily}
\SetAlCapFnt{\footnotesize\rmfamily}
\SetAlCapNameFnt{\footnotesize\rmfamily}
\SetAlTitleFnt{\footnotesize\rmfamily}

\SetCommentSty{AlgoCommentFont}

\SetNlSty{AlgoNlFont}{}{:}

\usepackage[hidelinks]{hyperref} 

\usepackage{textcomp}
\usepackage{verbatim}

\usepackage{stfloats} 

\usepackage{url} 

\addto\captionsenglish{}

\usepackage{balance}
\begin{document}
\title{Stochastic Design of Active RIS–Assisted Satellite Downlinks under Interference, Folded Noise, and EIRP Constraints}
\author{Muhammad Khalil, \textit{Member, IEEE}, Ke Wang, \textit{Senior Member, IEEE}, and Jinho Choi, \textit{Fellow, IEEE}
\thanks{Muhammad Khalil and Ke Wang are with the School of Engineering, RMIT University, Melbourne, Australia (e-mail: muhammad.khalil@rmit.edu.au; ke.wang@rmit.edu.au). Jinho Choi is with the University of Adelaide, Adelaide, Australia (e-mail: jinho.choi@adelaide.edu.au).}
}
\maketitle

\begin{abstract}
Active reconfigurable intelligent surfaces (RISs) can mitigate the double-fading loss of passive reflection in satellite downlinks, but their benefits are limited by random cochannel interference, gain-dependent amplifier noise, and regulatory emission constraints. This paper develops a stochastic reliability framework for active RIS--assisted satellite downlinks by modeling the desired and interfering channels, receiver noise, and RIS amplifier noise as random variables, leading to an instantaneous Signal-to-Interference-plus-Noise Ratio (SINR) representation that explicitly captures folded (cascaded) amplifier noise. The resulting model reveals a finite high-gain SINR ceiling.

To guarantee a target outage probability, we formulate a chance-constrained max–SINR design that jointly optimizes the binary RIS configuration and the common amplification gain. The chance constraint is handled via a sample-average approximation (SAA) with a violation budget, and the resulting feasibility problem is solved as a mixed-integer second-order cone program (MISOCP) within a bisection search over the SINR threshold. Practical implementability is ensured by restricting the gain to the admissible range imposed by small-signal stability and effective isotropic radiated power (EIRP) limits. In addition, we derive realization-wise SINR envelopes based on eigenvalue and $\ell_1$ bounds, which provide interpretable performance limits and fast diagnostics.

Monte Carlo results confirm that the envelopes tightly bracket the simulated SINR, reproduce the predicted saturation behavior, and quantify performance degradation as interference increases. These results provide a solver-ready, reliability-targeting design methodology whose achieved reliability is validated via Monte-Carlo testing for active RIS--assisted satellite downlinks under realistic randomness and hardware constraints.

\end{abstract}


\begin{table*}[t]
\centering
\renewcommand{\arraystretch}{1.25}
\setlength{\tabcolsep}{6pt}
\footnotesize
\caption{Main contributions of the paper.}
\label{tab:main-contributions}
\begin{tabularx}{\textwidth}{@{}l p{2.6cm} X@{}}
\toprule
\textbf{\#} & \textbf{Title} & \textbf{Key ideas \& significance} \\
\midrule

\textbf{C1} & \textbf{Stochastic SINR Bounds for Active RIS-Satellite Links} &
Derives simple closed-form SINR upper/lower bounds that hold for each random
channel realization, capturing fading, multi-satellite interference, and active-RIS amplification effects.\emph{Impact:} Enables fast reliability assessment without extensive Monte Carlo
simulation. \\

\textbf{C2} & \textbf{Reliability-Targeting Joint Design} &
Poses a chance-constrained joint design of discrete RIS phases and a continuous active-RIS gain targeting an SINR non-outage level $1-\epsilon$ under stochastic fading and interference, and solves an SAA--MISOCP approximation whose achieved reliability is validated via out-of-sample Monte-Carlo evaluation.
\emph{Impact:} Solver-ready reliability-targeting configuration under practical gain (stability/EIRP) limits. \\

\textbf{C3} & \textbf{Folded-Noise Modeling and SINR Ceiling Law} & Introduces a gain-dependent folded-noise model that captures how amplifier noise propagates through the RIS-assisted link, and proves a finite reliability-guaranteed SINR ceiling under stochastic interference, implying
that increasing $g$ beyond a threshold yields diminishing (or no) returns. \emph{Impact:} Explains why active-RIS amplification saturates and when the link becomes noise-limited. \\

\textbf{C4} & \textbf{Hardware- and Regulation-Compliant Design Constraints} & Integrates amplifier stability and EIRP regulations directly into the reliability-aware optimization via convex reformulations, ensuring
implementable solutions under physical hardware limits. \emph{Impact:} Enables certifiable active-RIS operation under practical stability/EIRP constraints in satellite downlinks. \\

\bottomrule
\end{tabularx}
\end{table*}

\section{Introduction}

Reconfigurable intelligent surfaces (RISs) have emerged as a practical way to shape wireless propagation and improve coverage in future networks \cite{Dajer2021}. By using electronically tunable reflecting elements, an RIS can coherently reconfigure incident waves and enhance the received signal without additional RF chains \cite{Li2023c}. Early studies showed that, under favorable geometry (e.g., blockage scenarios), the received power can scale roughly with the square of the number of RIS elements \cite{Basar2019}. Subsequent analyses incorporated fading and quantified performance via SINR/outage distributions, confirming the reliability improvements enabled by RIS-assisted virtual line-of-sight paths \cite{Ma2024}. These advances have motivated broad investigations of RISs for coverage extension, physical-layer security, and localization, with surveys emphasizing their spectral-and energy-efficiency potential for beyond-5G/6G systems \cite{Yildirim2021,Pan2021,Khalil2024,Liu2021}.

Motivated by these terrestrial results, RIS-assisted architectures are also being explored for non-terrestrial and satellite communications. Multibeam SatCom with aggressive frequency reuse offers high throughput and wide-area service; however, it is limited by severe path loss and co-channel interference due to beam overlap \cite{Zheng2024a}. Introducing an RIS into the satellite downlink can provide a controllable reflection path to mitigate shadowing and strengthen the received signal \cite{Toka2024}. For example, \cite{Kim2024} reports improved sum-rate and coverage in RIS-aided multibeam GEO systems, while \cite{Wang2021a} derives outage and rate expressions for an RIS-assisted LEO scenario using a UAV-mounted reflector, demonstrating gains over obstructed direct links. Overall, these works indicate that RISs can complement SatCom by extending coverage into deep fades and coverage holes.

Most existing RIS studies in both terrestrial and SatCom settings focus on \emph{passive} surfaces that only apply controllable phase shifts. Although power-efficient, passive RISs suffer from a \emph{multiplicative (double-fading)} limitation: the end-to-end gain is the product of the transmitter--RIS and RIS--receiver links, which can be especially weak when the two-hop path loss is large or a strong direct path exists \cite{Zhou2024}. This is critical for satellite downlinks, where long satellite--RISs and RIS--user distances compound attenuation, so passive RISs often provide modest SINR/rate gains unless very large apertures are used or the direct link is blocked \cite{Chen2021}. Empirical findings report only limited improvement in typical LoS terrestrial deployments \cite{Zhang2023}, and SatCom studies further note that passive reflections may also reradiate co-channel interference from adjacent beams, thereby reducing the net benefit under full-frequency reuse \cite{Zheng2025}.

To overcome these limitations, \emph{active} RIS architectures embed low-power amplifiers in each element to boost the reflected signal and partially compensate for the double-fading loss \cite{Wu2019a,Khalil2022}. Prototype results indicate that active reflection can surpass passive capacity ceilings by injecting power into the reradiated link \cite{Long2021}. However, amplification introduces new constraints: amplifier noise and distortion are folded through the cascaded channel, so increasing gain eventually yields diminishing SINR returns and a finite high-gain ceiling \cite{Khoshafa2021,Chen2023a,ObinnaNnamani2025,Yigit2025,Khalid2025}. In addition, practical operation must satisfy small-signal stability and regulatory EIRP limits, as excessive gain can trigger oscillations or violate emission constraints \cite{Dai2020,Gavriilidis2025}. Recent work has therefore studied active-RIS designs with joint beamforming/coefficients and architectures that limit noise accumulation and out-of-band radiation \cite{Zhu2022b,ZakkaElNashef2011,Wu2025,Zeng2022}, as well as security-oriented applications \cite{Lv2023}. In SatCom, active RIS concepts are still nascent; early studies suggest coverage gains in deep fades/NLoS (e.g., UAV-mounted active RISs) \cite{Niu2023,Tran2025}, but most analyses assume simplified links (often single-user or interference-free) and ideal gain control, leaving reliability-aware design under realistic multi-beam interference and stochastic fading largely open.

These observations motivate a unified design framework for active RIS--assisted satellite downlinks that jointly capture \emph{stochastic fading}, \emph{co-channel interference}, and \emph{hardware-induced noise}. Most optimization-driven RIS designs (passive or active) adopt deterministic links and idealized channel knowledge, targeting average or worst-case metrics \cite{Pan2022a}. In practice, satellite channels are highly time-varying due to user mobility, satellite motion, and atmospheric effects, while interference fluctuates with geometry and traffic load. Hence, guaranteeing a consistent quality-of-service requires a \emph{reliability-driven} formulation \cite{Kim2024a}. Chance-constrained methods are a standard tool for enforcing probabilistic SINR/outage guarantees in robust communications \cite{Zheng2010}, and initial RIS studies have incorporated outage behavior via SNR/SINR distributions and phase design \cite{Wang2021a}. However, a unified framework that \emph{jointly} optimizes discrete RIS states (e.g., binary phase shifts) and continuous active gains under probabilistic SINR constraints, while explicitly enforcing stability and EIRP limits, is still missing. This gap motivates our work.

In this paper, we develop a stochastic optimization framework for an active RIS--assisted satellite downlink under random fading and co-channel interference. The main contributions are as follows: (i) we establish a stochastic SINR model in which desired and interfering channels, as well as both receiver thermal noise and RIS amplifier noise, are treated as random variables; (ii) we pose a reliability-aware design via a chance constraint on the SINR and derive a tractable sample-average approximation (SAA) solved by a mixed-integer second-order cone program (MISOCP); (iii) we introduce a folded, gain-dependent noise model that explains SINR saturation at high gains; and (iv) we embed practical hardware and regulatory constraints—small--signal stability and EIRP limits—directly into the optimization, ensuring physically realizable designs. The major contributions are summarized in Table~\ref{tab:main-contributions}.

The remainder of this paper is organized as follows. Section~\ref{sec:2} presents the stochastic system model (Rician block fading, binary RIS reflection, and gain-dependent folded noise). Section~\ref{sec:3} formulates a chance-constrained reliability design and derives an SAA-based MISOCP to jointly optimize the RIS phases and gain. Section~\ref{sec:4} develops interference-aware SINR envelopes and the high-gain ceiling under amplifier noise. Section~\ref{sec:5} provides numerical validation under stability/EIRP-limited gain and explores reliability over $(N,M,g)$. Section~\ref{sec:6} concludes and outlines extensions to dual-band operation, data-driven RIS control, and measured channels.

\section{\label{sec:2}System Model (Stochastic)}

\begin{figure}[!t]
\centering
\includegraphics[width=\columnwidth,trim=2bp 0bp 0bp 0bp,clip]{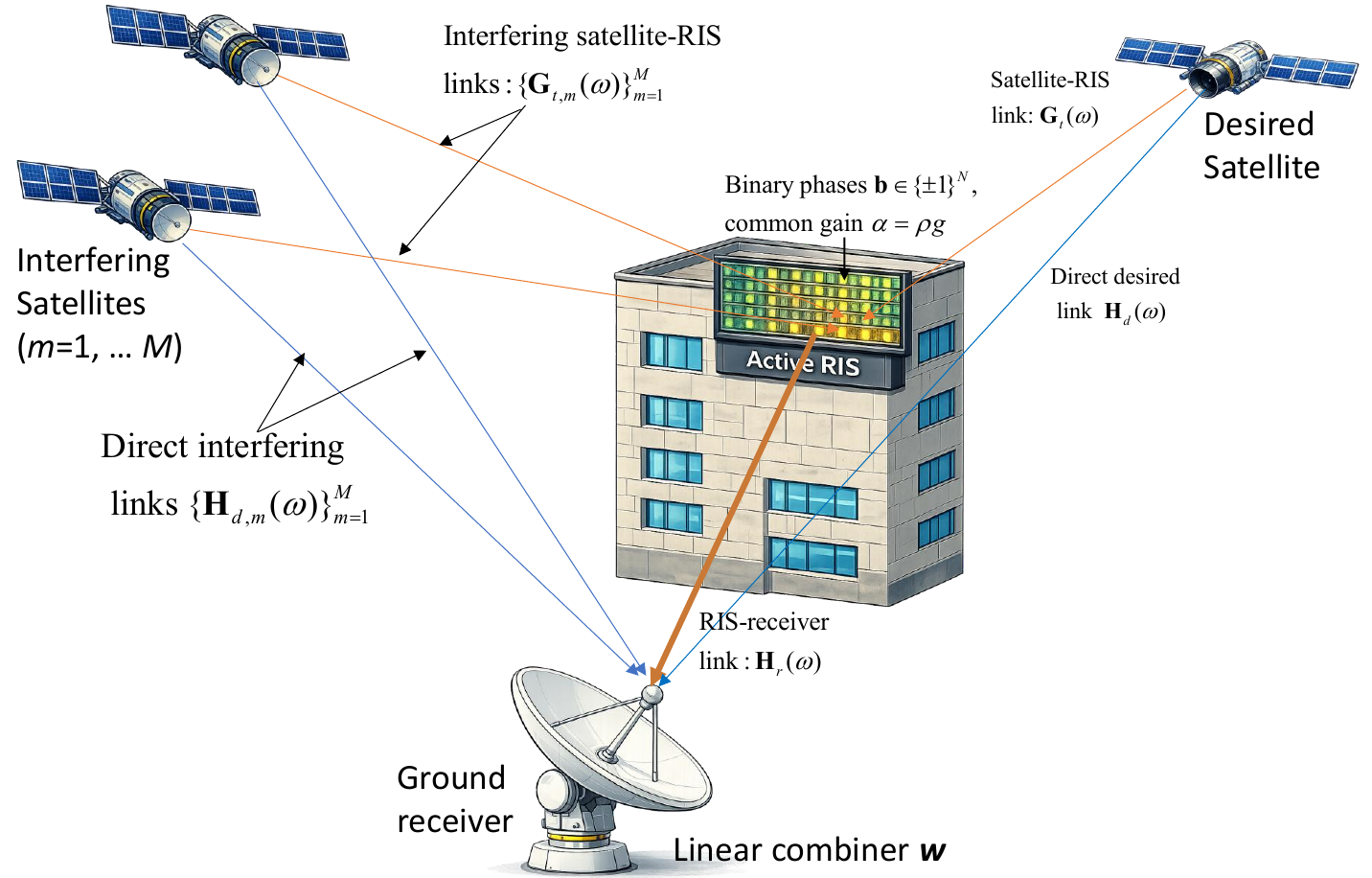}
\caption{\footnotesize Active RIS--assisted satellite downlink with one desired satellite and $M$ cochannel interferers. Desired and interfering signals reach the ground receiver via direct links $\mathbf H_d(\omega)$ and $\{\mathbf H_{d,m}(\omega)\}$ and via the RIS-assisted paths through $\mathbf G_t(\omega)$, $\{\mathbf G_{t,m}(\omega)\}$, and $\mathbf H_r(\omega)$. Blue links denote desired propagation; red links denote interference.}
\label{fig:system-model}
\end{figure}

We consider a narrowband satellite downlink assisted by an \emph{active} RIS, as illustrated in Fig.~\ref{fig:system-model}. One desired satellite transmits to a ground receiver, while $M$ neighboring satellites reuse the same carrier, creating cochannel interference. The satellite--RIS, RIS--receiver, and direct satellite--receiver channels follow a block-fading model: the channel coefficients remain constant within each coherence block and change independently from one block to the next. Throughout, the fading coefficients and noise terms are modeled as random variables.

Model components:
\begin{itemize}
\item \textbf{Topology:} one desired satellite, $M$ cochannel interferers, one active RIS with $N$ elements,
and a ground receiver with combiner $\mathbf w$ (Fig.~1).
\item \textbf{Channels:} block-fading Rician links (satellite--RIS, RIS--receiver, and satellite--receiver) \cite{Wang2021a,Kim2024}.
\item \textbf{Active RIS:} two-state phase control $\mathbf b\in\{\pm1\}^N$ and common gain $g$ \cite{Long2021,Wu2019a}.
\item \textbf{Noise:} receiver thermal noise and gain-dependent RIS amplifier noise folded through $\mathbf H_r(\cdot)$ \cite{Chen2023a,Khoshafa2021}.
\end{itemize}

The desired satellite employs a fixed precoder $\mathbf f\in\mathbb C^{N_t}$, and the ground receiver applies a linear combiner $\mathbf w\in\mathbb C^{N_r}$. The interferer $m$ uses $\mathbf f_m$. The transmitted symbols are modeled as zero-mean proper complex Gaussian:
\[
x\sim\mathcal{CN}(0,P_{\mathrm d}),\qquad
x_m\sim\mathcal{CN}(0,P_m),
\]
independent across transmitters and independent of all noise sources and channels.

Each hop (satellite$\,\to\,$RIS, RIS$\,\to\,$receiver, and satellite$\,\to\,$receiver) follows a Rician law with elevation-dependent path LoS and a Rician $K$-factor $K(\varepsilon)\ge 0$, defined as the ratio between the average power in the deterministic LoS component and the average power in the scattered NLoS component.
\begin{align}
\mathbf H(\omega)
&=\sqrt{\beta_{\mathrm{fs}}(s,\lambda)}\!
\left(
\sqrt{\tfrac{K(\varepsilon)}{K(\varepsilon)+1}}\mathbf H_{\!\mathrm{LoS}}
+\sqrt{\tfrac{1}{K(\varepsilon)+1}}\mathbf H_{\!\mathrm{NLoS}}(\omega)
\right),
\label{eq:rician-stoch}
\end{align}
where $\beta_{\mathrm{fs}}(s,\lambda)=(4\pi s/\lambda)^{-2}$ captures free-space path loss.
$\mathbf H_{\!\mathrm{NLoS}}(\omega)$ has i.i.d. $\mathcal{CN}(0,1)$ entries,
and $\mathbf H_{\!\mathrm{LoS}}$ is the deterministic array response.
For the desired link, we denote $\mathbf H_d(\omega)$ (direct path),
$\mathbf G_t(\omega)$ (satellite$\!\to\!$RIS), and $\mathbf H_r(\omega)$ (RIS$\!\to\!$receiver).
For the interferers $m$, we use $\mathbf H_{d,m}(\omega)$ and $\mathbf G_{t,m}(\omega)$.
All satellites illuminate the same RIS, so $\mathbf H_r(\omega)$ is common.

Each RIS element applies a binary phase reflection ($0$ or $\pi$) and a common small-signal gain $g$, scaled by the passive retention factor $\rho$. This captures a simple yet practical two-state phase control with a uniform amplifier setting across the array, consistent with first-generation active RIS prototypes~\cite{Long2021}.
Specifically,
\[
\Gamma_i=\alpha\,b_i,\qquad
\alpha=\rho\,g,\qquad
b_i\in\{+1,-1\},\ \ i=1,\ldots,N,
\]
where $0<\rho\le1$ models passive retention and $g\ge0$ is the active gain.
Let $\mathbf b=[b_1,\ldots,b_N]^{\mathsf T}$ and $\boldsymbol\Gamma=\alpha\,\mathrm{diag}(\mathbf b)$.

To calculate the cascaded coefficients, let $\mathbf h_{r,i}(\omega)$ denote the
$i$-th column of the RIS–receiver channel $\mathbf H_r(\omega)$, and
let $\mathbf g_{t,i}(\omega)$ denote the $i$-th row of the satellite–RIS channel
$\mathbf G_t(\omega)$.  
The corresponding random complex scalars are
\begin{align}
d(\omega)&=\mathbf w^{\mathsf H}\mathbf H_d(\omega)\mathbf f, &
u_i(\omega)&=\big(\mathbf w^{\mathsf H}\mathbf h_{r,i}(\omega)\big)
              \big(\mathbf g_{t,i}^{\mathsf T}(\omega)\mathbf f\big),
\end{align}
and the vector of RIS–mediated coefficients is
$\mathbf u(\omega)=[u_1(\omega),\ldots,u_N(\omega)]^{\mathsf T}$.  
For the $m$-th interferer, the same construction applies after replacing
$(\mathbf H_d,\mathbf G_t,\mathbf f)$ with $(\mathbf H_{d,m},\mathbf G_{t,m},\mathbf f_m)$, yielding $d_m(\omega)$
and $\mathbf u_m(\omega)$.

To account for the receiver and RIS noise, note that the receiver thermal disturbance is
$\mathbf n_r\!\sim\!\mathcal{CN}(\mathbf0,N_0\mathbf I)$.  
Each RIS element also introduces an independent or correlated amplifier noise term
$\mathbf n_{\mathrm{RA}}(\omega)\!\in\!\mathbb C^{N}$, modeled as a zero-mean proper complex Gaussian with covariance
\[
\mathbb E[\mathbf n_{\mathrm{RA}}(\omega)\mathbf n_{\mathrm{RA}}^{\mathsf H}(\omega)]
=\boldsymbol\Sigma_a(g,\omega)\succeq\mathbf0.
\]
A convenient small-signal law expresses the per-element variance as
$\sigma_a^2(g)=\sigma_{\min}^2+\eta g^2$, leading to
$\boldsymbol\Sigma_a(g,\omega)=\sigma_a^2(g)\mathbf R_a(\omega)$,
where $\mathbf R_a(\omega)=\mathbf I$ represents the i.i.d. case and the non-diagonal $\mathbf R_a(\omega)$ captures spatial correlation.  

After propagation through the RIS–receiver channel $\mathbf H_r(\omega)$ and combining by $\mathbf w$, the  total folded noise variance is
\begin{align}
\mathrm{Var}\!\big(\mathbf w^{\mathsf H}\mathbf H_r(\omega)\mathbf n_{\mathrm{RA}}(\omega)\big)
=\mathbf w^{\mathsf H}\mathbf H_r(\omega)
  \boldsymbol\Sigma_a(g,\omega)
  \mathbf H_r^{\mathsf H}(\omega)\mathbf w.
\label{eq:fold-general-stoch}
\end{align}
Here, \emph{folded noise} refers to the amplifier noise generated across the RIS elements, which propagates through $\mathbf H_r(\omega)$ and is spatially combined by $\mathbf w$, thereby \emph{folding} the spatial noise field at the RIS into an effective noise contribution at the receiver output.

For the common i.i.d. specialization,
\(
\boldsymbol\Sigma_a(g,\omega)=\sigma_a^2(g)\mathbf I
\)
with
\(
\sigma_a^2(g)=\sigma_{\min}^2+\eta g^2
\),
the folded-noise variance reduces to
\(
\sigma_a^2(g)L(\omega)
\),
where
\(
L(\omega)=\|\mathbf H_r^{\mathsf H}(\omega)\mathbf w\|_2^2
\)
quantifies the combiner--RIS channel alignment.

After accounting for the aforementioned channel and noise components, the scalar received signal during one coherence block can be expressed as follows:
\begin{align}
y(\omega)
&=\Big(d(\omega)+\rho g\,\mathbf u^{\mathsf T}(\omega)\mathbf b\Big)x
 +\sum_{m=1}^{M}\!\Big(d_m(\omega)+\rho g\,\mathbf u_m^{\mathsf T}(\omega)\mathbf b\Big)x_m
\nonumber\\
&\quad+\mathbf w^{\mathsf H}\mathbf n_r
 +\mathbf w^{\mathsf H}\mathbf H_r(\omega)\mathbf n_{\mathrm{RA}}(\omega).
\label{eq:y-stoch}
\end{align}
Here, the first term represents the desired signal reflected by the RIS, the summation accounts for cochannel interference from $M$ neighboring satellites, and the last two terms correspond to the receiver thermal noise and folded amplifier noise from the RIS, respectively.

To obtain a phase–aligned representation, let $\phi(\omega)=\arg(d(\omega))$ denote the instantaneous phase of the direct component. By rotating all reflected terms by $e^{-j\phi(\omega)}$, we define $\tilde{\mathbf u}(\omega)=\mathbf u(\omega)e^{-j\phi(\omega)}$ so that the desired component $d(\omega)e^{-j\phi(\omega)}$ becomes real and positive.   The resulting real-valued quantities are
\[
\mathbf r(\omega)=\Re\{\tilde{\mathbf u}(\omega)\},\qquad
\mathbf Q(\omega)\triangleq
\Re\!\left\{\tilde{\mathbf u}(\omega)\tilde{\mathbf u}^{\mathsf H}(\omega)\right\}.
\]
Since $\mathbf b\in\{\pm1\}^N\subset\mathbb R^N$ is real, only the symmetric real part contributes, and
\[
\mathbf b^{\mathsf T}\mathbf Q(\omega)\mathbf b
=\left|\tilde{\mathbf u}^{\mathsf T}(\omega)\mathbf b\right|^2\ge0.
\]
Hence, $\mathbf Q(\omega)$ is positive semidefinite in the induced real quadratic form used throughout.

Analogous definitions apply to each interferer $m$, using $\phi_m(\omega)=\arg(d_m(\omega))$. This rotation simplifies subsequent derivations since all cross terms between the direct and reflected signals become real-valued, and the SINR expression can be presented in a purely real form without loss of generality.

To express the received signal in a compact quadratic form, define the following real-valued coefficients for each random realization~$\omega$:
\begin{gather}
A(\omega)=|d(\omega)|^{2},B(\mathbf{b},\omega)=2\rho|d(\omega)|\,\mathbf{b}^{\mathsf{T}}\mathbf{r}(\omega),C(\mathbf{b},\omega)\nonumber \\
=\rho^{2}\mathbf{b}^{\mathsf{T}}\mathbf{Q}(\omega)\mathbf{b}.\label{eq:def-ABC-stoch}
\end{gather}
Analogous quantities $A_m(\omega)$, $B_m(\mathbf b,\omega)$, and $C_m(\mathbf b,\omega)$ are defined for each interferer $m$ as follows: 
\begin{gather}
A_{m}(\omega)=|d_{m}(\omega)|^{2},\quad
B_{m}(\mathbf{b},\omega)=2\rho|d_{m}(\omega)|\,\mathbf{b}^{\mathsf{T}}\mathbf{r}_{m}(\omega)\nonumber \\
C_{m}(\mathbf{b},\omega)=\rho^{2}\,\mathbf{b}^{\mathsf{T}}\mathbf{Q}_{m}(\omega)\mathbf{b}.
\label{eq:def-ABCm-stoch}
\end{gather}

The noise terms introduced earlier on the receiver-side provide the random coefficients
\[
D_0(\omega)=N_0\|\mathbf w\|_2^2+\sigma_{\min}^2L(\omega),\qquad
D_1(\omega)=\eta L(\omega),
\]
where $L(\omega)=\|\mathbf H_r^{\mathsf H}(\omega)\mathbf w\|_2^{2}$ is the folded gain factor that links the RIS to the combiner.  

Combining all components, the instantaneous SINR  for a given binary configuration~$\mathbf b$ and amplification gain~$g$ is expressed compactly as
\begin{align}
S(\mathbf b,g;\omega) &\triangleq A(\omega)+gB(\mathbf b,\omega)+g^2C(\mathbf b,\omega),\\
I(\mathbf b,g;\omega) &\triangleq \sum_{m=1}^{M}P_m\!\left(A_m(\omega)+gB_m(\mathbf b,\omega)+g^2C_m(\mathbf b,\omega)\right),\\
\mathrm{SINR}(\mathbf b,g;\omega)
&= \frac{P_{\mathrm d}\,S(\mathbf b,g;\omega)}{D_0(\omega)+D_1(\omega)g^2+I(\mathbf b,g;\omega)}
\label{eq:sinr-compact-stoch}.
\end{align}

Equation~\eqref{eq:sinr-compact-stoch} follows by expanding the squared magnitudes of the effective channels in \eqref{eq:y-stoch} to obtain the quadratic forms in \eqref{eq:def-ABC-stoch}--\eqref{eq:def-ABCm-stoch}, and by using the
folded-noise variance expression in \eqref{eq:fold-general-stoch} (and its i.i.d.\ specialization).
Detailed algebraic derivations are available in Appendix A.

Substituting the above definitions yields an SINR expression in quadratic form, which explicitly links the binary phase configuration $\mathbf b$ and the active gain $g$ to stochastic fading, cochannel interference, and gain-dependent noise. This compact representation serves as the basis for the chance-constrained optimization developed in Sec.~\ref{sec:3}.

The amplification factor $g$ is subject to both circuit stability and regulatory
emission limits.  From a small–signal perspective, each RIS element behaves as a
two–port device; satisfying the Rollett stability conditions ($K_{\mathrm R}>1$, $|\Delta|<1$) imposes an upper bound on the realizable gain:
\[
0\le g\le g_{\mathrm{stab}}\triangleq\mu\,\mathrm{MAG},
\]
where $\mathrm{MAG}$ denotes the maximum available gain of the cell, and
$\mu\!\in\!(0,1)$ is a chosen safety factor that maintains stability margins under
manufacturing or temperature variations.
In addition to stability, the effective isotropic radiated power (EIRP) from each RIS cell must satisfy regulatory emission constraints. With equal-gain two-state reflection (\(\alpha=\rho g\)), the instantaneous reradiated power of element \(i\) is \(\rho^{2}g^{2}\Psi_{i}(\omega)\), where the random incident power in one coherence block is
\[
\Psi_{i}(\omega)
=
P_{\mathrm d}\big|\mathbf g_{t,i}^{\mathsf T}(\omega)\mathbf f\big|^{2}
+\sum_{m=1}^{M}P_{m}\big|\mathbf g_{t,m,i}^{\mathsf T}(\omega)\mathbf f_{m}\big|^{2}.
\]
Therefore, enforcing the per-cell emission limit \(P_{\mathrm{cell}}^{\max}\) requires
\[
\rho^{2}g^{2}\Psi_{i}(\omega)\le P_{\mathrm{cell}}^{\max},
\qquad \forall i,\ \forall \omega.
\]
Because \(\Psi_{i}(\omega)\) is random (through stochastic propagation and interference), the gain constraint on \(g\) is imposed using one of the following statistically conservative formulations. Define the worst-cell incident power as
\[
\Psi_{\max}(\omega)\triangleq \max_{i=1,\ldots,N}\Psi_i(\omega).
\]

\smallskip
\textit{1) Sample-wise worst-case (scenario-robust) bound:}
Using the same $S$ samples $\{\omega_s\}_{s=1}^{S}$ as in the SAA, enforce the EIRP constraint for every sampled realization:
\begin{equation}
g_{\mathrm{EIRP,wc}}^{(S)}
=\frac{\sqrt{P_{\mathrm{cell}}^{\max}}}
{\rho\,\sqrt{\max_{s=1,\ldots,S}\Psi_{\max}(\omega_s)}}.
\label{eq:gEIRP-wc}
\end{equation}

\textit{2) A quantile (chance-constrained) bound:} an outage level
$\alpha\!\in\!(0,1)$ is permitted such that
$\Pr\{\rho^{2}g^{2}\Psi_{\max}(\omega)\le P_{\mathrm{cell}}^{\max}\}\ge1-\alpha$.
Let $q_{1-\alpha}$ denote the $(1-\alpha)$ quantile of $\Psi_{\max}(\omega)$;
then
\begin{equation}
g_{\mathrm{EIRP},\,1-\alpha}=\frac{\sqrt{P_{\mathrm{cell}}^{\max}}}{\rho\,\sqrt{q_{1-\alpha}}}.\label{eq:gEIRP-quantile}
\end{equation}

\textit{3) Moment-based (Cantelli) bound:} when only the first two moments of
$\Psi_{\max}(\omega)$ are known, Cantelli’s inequality provides a closed-form, safe bound.
\begin{equation}
g_{\mathrm{EIRP,Cantelli}}=\frac{\sqrt{P_{\mathrm{cell}}^{\max}}}{\rho\,\sqrt{\mu_{\max}+c_{\alpha}\sigma_{\max}}},\qquad c_{\alpha}=\sqrt{\tfrac{1-\alpha}{\alpha}}.\label{eq:gEIRP-cantelli}
\end{equation}
where $\mu_{\max}$ and $\sigma_{\max}$ are the means and standard deviations of
$\Psi_{\max}(\omega)$, respectively.

\smallskip
The admissible gain must satisfy both the small-signal stability requirement and
the selected EIRP constraint, leading to the overall range
\begin{equation}
0\le g\le g_{\max}=\min\{\,g_{\mathrm{stab}},\,g_{\mathrm{EIRP}}\,\}.\label{eq:gmax-stochastic}
\end{equation}
where $g_{\mathrm{EIRP}}$ corresponds to one of the bounds above, chosen according to the desired reliability level.

This system model unifies random satellite geometry, cochannel interference, and hardware-imposed gain limits within a single stochastic framework. In contrast to deterministic or average-link formulations, \eqref{eq:sinr-compact-stoch} makes explicit how random fading, gain-dependent RIS amplifier noise, and interference jointly determine the instantaneous SINR over each coherence block, while \eqref{eq:gmax-stochastic} enforces physically admissible amplification through stability and EIRP constraints. This stochastic representation enables the reliability-aware, chance-constrained optimization developed in Sec.~\ref{sec:3} and provides a physically consistent foundation for the SAA--MISOCP formulation.

\paragraph*{Folded RIS--Amplifier Noise (Stochastic).}
\label{sec:folded-noise}
We now characterize how RIS amplifier noise propagates through the cascaded channel and determines the stochastic coefficients $D_0(\omega)$ and $D_1(\omega)$ in the SINR denominator.  This connection is essential because these coefficients encapsulate the total noise power observed at the receiver, thereby determining the achievable reliability for any feasible RIS gain $g$. By quantifying how spatial correlation and gain-dependent amplification modify the effective noise, this section provides a statistical link between the RIS hardware behaviour and the stochastic optimization model developed earlier.

We model the per-element amplifier noise as a zero-mean proper complex vector $\mathbf n_{\mathrm{RA}}(\omega)\!\in\!\mathbb C^N$ with covariance
\begin{align}
\mathbb E[\mathbf n_{\mathrm{RA}}(\omega)]&=\mathbf 0, &
\mathbb E[\mathbf n_{\mathrm{RA}}(\omega)\mathbf n_{\mathrm{RA}}^{\mathsf H}(\omega)]
&=\boldsymbol\Sigma_a(g,\omega)\succeq\mathbf 0,
\label{eq:sigma-general}
\end{align}
independent of the transmitted data symbols $\{x,x_m\}$ and the receiver noise $\mathbf n_r$.

Before proceeding, we briefly outline the structure of this subsection to clarify the modeling logic.  We first characterize how the internal amplifier noise scales with the RIS gain~$g$, which forms the basis of the gain-dependent covariance model.  Next, we analyze how this noise component is propagated through the RIS–receiver channel and folded at the combiner output, leading to the expressions for the random coefficients $D_0(\omega)$ and $D_1(\omega)$ in the SINR denominator.  Finally, we establish analytical bounds and special cases that connect this general stochastic model to the commonly used i.i.d. and diagonal noise
assumptions.

\paragraph*{Gain dependence.}
The internal noise generated by RIS amplifiers increases with operating gain. We model the amplifier-noise covariance as
\begin{align}
\boldsymbol\Sigma_a(g,\omega)
= \boldsymbol\Sigma_{\min}(\omega)+g^2\boldsymbol\Sigma_{\mathrm{ex}}(\omega),
\label{eq:sigma-affine}
\end{align}
where \(\boldsymbol\Sigma_{\min}(\omega)\succeq\mathbf0\) is the gain-independent baseline covariance and
\(\boldsymbol\Sigma_{\mathrm{ex}}(\omega)\succeq\mathbf0\) is the gain-dependent excess covariance. This PSD affine model covers i.i.d., diagonal, and correlated amplifier-noise structures in a unified form.

Unless stated otherwise, simulations and the SAA formulation use the i.i.d. specialization
\(\boldsymbol\Sigma_{\min}(\omega)=\sigma_{\min}^2\mathbf I\) and
\(\boldsymbol\Sigma_{\mathrm{ex}}(\omega)=\eta\mathbf I\), so that
\(\boldsymbol\Sigma_a(g,\omega)=(\sigma_{\min}^2+\eta g^2)\mathbf I\).
Under this specialization, \eqref{eq:D0D1-stoch} reduces to the scalar coefficients used in \eqref{eq:sinr-compact-stoch} and Table~\ref{tab:saa-coeff}.

\paragraph*{Folded variance at the receiver.}
After propagation through the RIS–receiver channel $\mathbf H_r(\omega)$ and
combining by $\mathbf w$, the amplifier noise folds to the output with variance
\begin{align}
\mathrm{Var}_\omega\!\big(\mathbf w^{\mathsf H}\mathbf H_r(\omega)\mathbf n_{\mathrm{RA}}(\omega)\big)
=\mathbf w^{\mathsf H}\mathbf H_r(\omega)\,
  \boldsymbol\Sigma_a(g,\omega)\,
  \mathbf H_r^{\mathsf H}(\omega)\mathbf w.
\label{eq:fold-var}
\end{align}
Using \eqref{eq:sigma-affine}, this separates into a gain-independent part and a gain-dependent part:
\begin{align}
D_{0,\mathrm{RA}}(\omega)
&=\mathbf w^{\mathsf H}\mathbf H_r(\omega)\,\boldsymbol\Sigma_{\min}(\omega)\,\mathbf H_r^{\mathsf H}(\omega)\mathbf w,\\
D_{1,\mathrm{RA}}(\omega)
&=\mathbf w^{\mathsf H}\mathbf H_r(\omega)\,\boldsymbol\Sigma_{\mathrm{ex}}(\omega)\,\mathbf H_r^{\mathsf H}(\omega)\mathbf w.
\end{align}
Adding the receiver thermal noise yields the random denominator coefficients
\begin{align}
D_0(\omega)&=N_0\|\mathbf w\|_2^2+D_{0,\mathrm{RA}}(\omega), &
D_1(\omega)&=D_{1,\mathrm{RA}}(\omega),
\label{eq:D0D1-stoch}
\end{align} so the total noise term in \eqref{eq:sinr-compact-stoch} is $D_0(\omega)+g^2D_1(\omega)$.

\paragraph*{Geometric interpretation (Rayleigh–Ritz bounds).}
To gain intuition about how the RIS channel geometry influences the folded
amplifier noise, we employ a Rayleigh–Ritz characterization.
This approach expresses each quadratic form
$\mathbf w^{\mathsf H}\mathbf H_r(\omega)\mathbf A\mathbf H_r^{\mathsf H}(\omega)\mathbf w$
in terms of its principal eigenvalues and the effective channel alignment of the
receive combiner.  
Specifically, let
\[
L(\omega)\triangleq\|\mathbf H_r^{\mathsf H}(\omega)\mathbf w\|_2^2
=\mathbf w^{\mathsf H}\mathbf H_r(\omega)\mathbf H_r^{\mathsf H}(\omega)\mathbf w
\] represent the squared channel projection between the RIS and the receiver.  
For any positive semidefinite matrix $\mathbf A\succeq\mathbf 0$, the
Rayleigh–Ritz inequality gives
\[
\lambda_{\min}(\mathbf A)\,L(\omega)\ \le\
\mathbf w^{\mathsf H}\mathbf H_r(\omega)\mathbf A\mathbf H_r^{\mathsf H}(\omega)\mathbf w
\ \le\ \lambda_{\max}(\mathbf A)\,L(\omega).
\]
Applying this relation to the covariance matrices
$\boldsymbol\Sigma_{\min}(\omega)$ and $\boldsymbol\Sigma_{\mathrm{ex}}(\omega)$
yields the corresponding bounds on the folded noise terms:
\begin{align}
\lambda_{\min}\!\big(\boldsymbol\Sigma_{\min}(\omega)\big)L(\omega)
\ \le\ D_{0,\mathrm{RA}}(\omega)\ \le\
\lambda_{\max}\!\big(\boldsymbol\Sigma_{\min}(\omega)\big)L(\omega),\nonumber\\
\lambda_{\min}\!\big(\boldsymbol\Sigma_{\mathrm{ex}}(\omega)\big)L(\omega)
\ \le\ D_{1}(\omega)\ \le\
\lambda_{\max}\!\big(\boldsymbol\Sigma_{\mathrm{ex}}(\omega)\big)L(\omega).
\label{eq:ritz-bounds}
\end{align}
These inequalities show that the scalar $L(\omega)$ quantifies the effective
alignment between the combiner and the RIS channel, while the eigenvalue spreads of $\boldsymbol\Sigma_{\min}(\omega)$ and $\boldsymbol\Sigma_{\mathrm{ex}}(\omega)$ determine how hardware noise correlation scales the baseline and gain-dependent components.

\paragraph*{Special cases.}
The Rayleigh–Ritz relations in \eqref{eq:ritz-bounds} provide general bounds
that apply to any positive semidefinite covariance matrices $\boldsymbol\Sigma_{\min}(\omega)$ and $\boldsymbol\Sigma_{\mathrm{ex}}(\omega)$. To interpret these results more concretely and connect the stochastic model to practical RIS implementations, we now examine several representative covariance structures.  Each case corresponds to a different level of spatial correlation among RIS amplifiers and illustrates how the folded noise coefficients $D_0(\omega)$ and $D_1(\omega)$ simplify under specific assumptions.

\smallskip
\noindent\textbf{1) i.i.d. equal-variance model:}
When all RIS amplifiers produce independent and identically distributed noise
with equal power, the covariance matrices reduce to
$\boldsymbol\Sigma_{\min}(\omega)=\sigma_{\min}^2\mathbf I$ and
$\boldsymbol\Sigma_{\mathrm{ex}}(\omega)=\eta\,\mathbf I$.  
Substituting these forms into \eqref{eq:D0D1-stoch} gives
\[
D_0(\omega)=N_0\|\mathbf w\|_2^2+\sigma_{\min}^2L(\omega),\qquad
D_1(\omega)=\eta\,L(\omega),
\]
which reproduces the familiar scalar law widely used in deterministic analyses. In this setting, all RIS elements contribute equally to the folded noise.

\smallskip
\noindent\textbf{2) Element-dependent, uncorrelated model:}
If each RIS element has a distinct amplifier with a different noise variance but remains uncorrelated with the others, the covariance matrices become diagonal:
\[
\boldsymbol\Sigma_{\min}(\omega)=\mathrm{diag}\{\sigma_{\min,1}^2,\ldots,\sigma_{\min,N}^2\},\quad
\boldsymbol\Sigma_{\mathrm{ex}}(\omega)=\mathrm{diag}\{\eta_1,\ldots,\eta_N\}.
\]
In this case, \begin{gather*}
D_{0,\mathrm{RA}}(\omega)=\sum_{i=1}^{N}\sigma_{\min,i}^{2}(\omega)\,|\mathbf{w}^{\mathsf{H}}\mathbf{h}_{r,i}(\omega)|^{2},\\
\:D_{1}(\omega)=\sum_{i=1}^{N}\eta_{i}(\omega)\,|\mathbf{w}^{\mathsf{H}}\mathbf{h}_{r,i}(\omega)|^{2},
\end{gather*}
where $\mathbf h_{r,i}(\omega)$ denotes the $i$-th column of $\mathbf H_r(\omega)$.  
This configuration reveals how spatially varying amplifier quality weights each
RIS path according to its coupling with the receiver.

\smallskip
\noindent\textbf{3) Correlated amplifier noise:}
In practical RIS hardware, mutual coupling and shared bias networks often induce
correlations among the amplifiers.  
The general formulation \eqref{eq:fold-var}–\eqref{eq:ritz-bounds} then applies
with arbitrary positive-semidefinite
$\boldsymbol\Sigma_{\min}(\omega)$ and $\boldsymbol\Sigma_{\mathrm{ex}}(\omega)$.  
This case captures spatially structured noise fields and allows for the analysis of how
correlation among RIS elements modifies the effective noise floor, potentially
tightening or relaxing the SINR constraints depending on the dominant
eigenmodes of the covariance matrices.

\paragraph*{Sample use in SAA.}
The quantities $D_0(\omega)$ and $D_1(\omega)$ derived above enter directly into the chance-constrained formulation of the optimization problem (Sec.~\ref{sec:3}).  
For each sampled realization $\omega_s$ of the random environment, these
quantities are evaluated as follows:
\[
D_{0,s}=D_0(\omega_s),\qquad D_{1,s}=D_1(\omega_s),
\]
using the definitions in \eqref{eq:D0D1-stoch}.  
In this way, every scenario in the sample-average approximation (SAA) inherits a distinct noise realization that reflects the specific RIS geometry, receiver combiner, and amplifier correlation structure of that sample.

Under standard scenario-sampling assumptions---i.i.d. samples $\{\omega_s\}_{s=1}^{S}$, finite moments of the random coefficients (e.g., $\mathbb{E}\|\mathbf H_r(\omega)\|_F^2<\infty$), and uniformly bounded eigenvalues of $\boldsymbol\Sigma_{\min}(\omega)$ and $\boldsymbol\Sigma_{\mathrm{ex}}(\omega)$---the sample averages converge to their population values as $S\to\infty$. Therefore, the SAA constraints become an increasingly accurate \emph{empirical} approximation of the original chance-constrained problem.

Because the resulting program is mixed-integer and relies on Big-$M$ activation, McCormick envelopes, and SOC reformulations, feasibility at finite $S$ should be interpreted as an empirical guarantee on the training scenarios, not as an exact probabilistic certificate. To evaluate reliability, we test the returned design on an independent out-of-sample Monte Carlo set $\{\tilde{\omega}_r\}_{r=1}^{S_{\mathrm{test}}}$ and report
\[
\widehat p_{\mathrm{succ}}
=\frac{1}{S_{\mathrm{test}}}\sum_{r=1}^{S_{\mathrm{test}}}
\mathbf 1\!\left\{\mathrm{SINR}(\mathbf b^{\star},g^{\star};\tilde{\omega}_r)\ge \tau^{\star}\right\},
\]
optionally with a binomial confidence interval.

These properties confirm that the stochastic formulation of $D_0(\omega)$ and $D_1(\omega)$ provides a physically consistent basis for the chance-constrained design developed in Sec.~\ref{sec:3}.

Overall, the resulting SAA--MISOCP solution $(\mathbf b^\star,g^\star,\tau^\star)$ bridges physical randomness (fading, interference, and folded amplifier noise) and reliability-driven system design by targeting the requirement
$\Pr\{\mathrm{SINR}(\mathbf b^\star,g^\star;\omega)\ge\tau^\star\}\ge 1-\varepsilon$
(up to the sampling accuracy of the SAA), while respecting the admissible gain interval $0\le g\le g_{\max}$ imposed by stability and EIRP limits in \eqref{eq:gmax-stochastic}. From an engineering viewpoint, this yields a quantifiable reliability--gain trade-off: increasing $g$ can strengthen the RIS-assisted component and improve coverage, but it also increases the gain-dependent folded-noise contribution through $D_1(\omega)$ and reduces feasibility margins under interference and emission constraints; conversely, smaller gains mitigate self-induced noise amplification and typically improve robustness across stochastic realizations. This interpretation motivates the ceiling analysis and envelope bounds in Sec.~\ref{sec:4}, which clarify when active amplification is beneficial and when noise- and interference-limited saturation dominates.



\section{\label{sec:3}Stochastic Problem Formulation and SAA--MISOCP}

In realistic satellite downlinks, fading and cochannel interference vary randomly across coherence blocks, so deterministic optimization cannot guarantee consistent link quality. To enforce reliability, we require the instantaneous SINR to exceed a target threshold $\tau$ with high probability $1-\varepsilon$, where $\varepsilon\in(0,1)$ denotes the allowable outage level (e.g., $\varepsilon=0.1$ corresponds to $90\%$ reliability). The design objective is therefore to determine the binary RIS configuration $\mathbf b$ and the amplifier gain $g$ that maximize this guaranteed SINR level.

With all channels being random, the coefficients $A,B,C,A_m,B_m,C_m,D_0,D_1$ in \eqref{eq:sinr-compact-stoch} become random functions of $\omega$. For any fixed $(\mathbf b,g)$, $\mathrm{SINR}(\mathbf b,g;\omega)$ is a random variable.

To ensure reliability under random channel variations, the design aims to maximize the guaranteed SINR threshold~$\tau$ subject to an allowable outage probability~$\varepsilon\!\in\!(0,1)$.   Specifically, the objective is to find the binary RIS configuration~$\mathbf b$ and the gain~$g$ that achieve
\begin{align}
\max_{\mathbf{b}\in\{\pm1\}^{N},\;0\le g\le g_{\max}}\ \tau\nonumber \\
\quad\text{s.t.}\quad\Pr\!\{\mathrm{SINR}(\mathbf{b},g;\omega)\ge\tau\}\ge1-\varepsilon.\label{eq:chance-prob}
\end{align}
This optimization ensures that, with probability at least~$1-\varepsilon$, the instantaneous SINR exceeds the target~$\tau$ over the random channel realizations.  Although problem~\eqref{eq:chance-prob} is generally non-convex and analytically intractable, it can be effectively approximated using the sample–average approximation (SAA) framework, which provides empirical probabilistic guarantees when the number of samples is sufficiently large.

To construct the SAA, we first generate $S$ independent and identically
distributed (i.i.d.) channel realizations
$\{\omega_s\}_{s=1}^{S}$ that represent possible random environments, including fading, elevation angles, and noise realizations.  
For each scenario~$s$, 
the stochastic coefficients
$\{A_s,B_s(\mathbf b),C_s(\mathbf b),A_{m,s},B_{m,s}(\mathbf b),C_{m,s}(\mathbf b),D_{0,s},D_{1,s}\}$
are obtained by evaluating the definitions in Sec.~\ref{sec:2} at the sampled realization
$\omega=\omega_s$. For convenience, Table~\ref{tab:saa-coeff} summarizes the
per-sample coefficients used in the SAA constraints.

\begin{table}[!t]
\centering
\footnotesize
\setlength{\tabcolsep}{5pt}
\renewcommand{\arraystretch}{1.15}
\caption{Per-sample stochastic coefficients for SAA (scenario $\omega_s$).}
\label{tab:saa-coeff}
\begin{tabularx}{\columnwidth}{@{}l X@{}}
\toprule
\textbf{Coefficient} & \textbf{Definition at $\omega=\omega_s$} \\
\midrule
$A_s$ & $|d(\omega_s)|^2$ \\

$B_s(\mathbf b)$ & $2\rho\,|d(\omega_s)|\,\mathbf b^{\mathsf T}\mathbf r(\omega_s)$ \\

$C_s(\mathbf b)$ & $\rho^2\,\mathbf b^{\mathsf T}\mathbf Q(\omega_s)\mathbf b$ \\

$A_{m,s}$ & $|d_m(\omega_s)|^2$ \\

$B_{m,s}(\mathbf b)$ & $2\rho\,|d_m(\omega_s)|\,\mathbf b^{\mathsf T}\mathbf r_m(\omega_s)$ \\

$C_{m,s}(\mathbf b)$ & $\rho^2\,\mathbf b^{\mathsf T}\mathbf Q_m(\omega_s)\mathbf b$ \\

$L(\omega_s)$ & $\|\mathbf H_r^{\mathsf H}(\omega_s)\mathbf w\|_2^2$ \\

$D_{0,s}$ & $N_0\|\mathbf w\|_2^2+\sigma_{\min}^2\,L(\omega_s)$ \\

$D_{1,s}$ & $\eta\,L(\omega_s)$ \\
\bottomrule
\end{tabularx}
\end{table}

The folding factor $L(\omega_s)$ quantifies how the RIS--receiver channel and the combiner
project the spatial amplifier-noise field into the receiver output. For each sample,
these coefficients fully determine the instantaneous SINR under \eqref{eq:sinr-compact-stoch}
and serve as deterministic inputs to the SAA optimization.


At a given SINR threshold~$\tau$, the per–sample feasibility condition becomes
\begin{gather}
P_{\mathrm d}\big(A_{s}+g\,B_{s}(\mathbf{b})+g^{2}C_{s}(\mathbf{b})\big)\ \ge\ \tau\!\nonumber \\
\left(D_{0,s}+D_{1,s}g^{2}+\sum_{m=1}^{M}P_{m}\big(A_{m,s}+g\,B_{m,s}(\mathbf{b})+g^{2}C_{m,s}(\mathbf{b})\big)\right).\label{eq:feas-int-stoch}
\end{gather}

This inequality ensures that the instantaneous SINR requirement is met for the $s$–th scenario. To impose the overall chance constraint $\Pr\{\cdot\}\ge1-\varepsilon$, no more than $\kappa=\lfloor\varepsilon S\rfloor$ samples are permitted to violate it. Binary indicator variables $v_s\!\in\!\{0,1\}$ mark these potential violations, leading to the joint conditions
\begin{equation}
\underbrace{\sum_{s=1}^{S}v_{s}\le\kappa}_{\text{violation budget}},\;
\underbrace{\mathcal{E}_{s}(\mathbf{y},\mathbf{s},\mathbf{u},g,t,\mathbf{z},\mathbf{Z};\tau)\ge-\Mbig v_{s}}_{\text{Big-\ensuremath{M} activation}},\;
s=1,\ldots,S,
\label{eq:bigM}
\end{equation}
where $\mathcal E_s(\cdot;\tau)$ denotes the linearized form of \eqref{eq:feas-int-stoch}. 
In the bisection oracle, $\Mbig$ is evaluated (or safely upper-bounded) for each queried value of $\tau$.


The Big-M constant is chosen as a conservative upper bound on the worst-case violation of $\mathcal E_s$ over the lifted feasible set (equivalently, over $t\in[0,g_{\max}^2]$):
\begin{equation}
\mathsf{M}_{\mathrm{big}}
=(1+\eta_M)\max_{s=1,\ldots,S}\!\left\{\tau\,\overline{R}_s-\underline{L}_s\right\},
\quad \eta_M\in[0.01,0.05],
\label{eq:Mbig-choice}
\end{equation}
where $\underline{L}_s$ and $\overline{R}_s$ are valid bounds satisfying
\begin{align}
\underline{L}_s
&\le
\inf_{\mathbf x\in\mathcal X_{\mathrm{lift}}}
P_{\mathrm d}\,\mathcal T_s(\mathbf x;\omega_s),\\
\overline{R}_s
&\ge
\sup_{\mathbf x\in\mathcal X_{\mathrm{lift}}}
\left[
D_{0,s}+D_{1,s}t+\sum_{m=1}^{M}P_m\,\mathcal T_{m,s}(\mathbf x;\omega_s)
\right],
\end{align}
with
\[
\mathbf x \triangleq (\mathbf y,\mathbf s,\mathbf u,g,t,\mathbf z,\mathbf Z),
\]
and $\mathcal X_{\mathrm{lift}}$ denoting the lifted feasible set defined by variable bounds, binary restrictions, McCormick envelopes, and SOC/linking constraints.
Accordingly, $v_s=0$ enforces $\mathcal E_s\ge0$, while $v_s=1$ allows relaxation via $-\mathsf{M}_{\mathrm{big}}v_s$, subject to $\sum_{s=1}^{S}v_s\le\kappa$.

\smallskip
Let $\mathbf{1}\in\mathbb{R}^{N}$ denote the length-$N$ all-ones vector. For each sample~$s$, define the auxiliary terms
\[
\mathbf q_s=\mathbf Q(\omega_s)\mathbf 1,\qquad
q_{0,s}=\mathbf 1^{\mathsf T}\mathbf Q(\omega_s)\mathbf 1,
\]
and, analogously, $\mathbf q_{m,s}$ and $q_{0,m,s}$ for each interferer $m$.
Using the binary mapping $b_i=2y_i-1$ and introducing the shared auxiliary
variables $(\mathbf s,\mathbf u,g,t,\mathbf z,\mathbf Z)$, the nonlinear
feasibility condition~\eqref{eq:feas-int-stoch} becomes \emph{linear} in all decision variables with sample-dependent coefficients:
\begin{equation}
\mathcal{E}_{s}(\cdot;\tau)\ \triangleq\
P_{\mathrm d}\,\mathcal{T}_{s}(\omega_{s})
-\tau\!\left(D_{0,s}+D_{1,s}t+\sum_{m=1}^{M}P_{m}\,\mathcal{T}_{m,s}(\omega_{s})\right).
\label{eq:lin-ineq-sample}
\end{equation}
\noindent where
\begin{IEEEeqnarray*}{rCl}
\mathcal{T}_{s}(\omega_{s}) \triangleq
& & A_{s}
-2\rho|d(\omega_{s})|\,g\,\mathbf{1}^{\mathsf{T}}\mathbf{r}(\omega_{s})
+4\rho|d(\omega_{s})|\,\mathbf{r}^{\mathsf{T}}(\omega_{s})\mathbf{u}
\nonumber\\
& &\quad
+\rho^{2}q_{0,s}\,t
-4\rho^{2}\mathbf{q}_{s}^{\mathsf{T}}\mathbf{z}
+4\rho^{2}\sum_{i,j}Q_{s,ij}Z_{ij},
\\[0.25em]
\mathcal{T}_{m,s}(\omega_{s}) \triangleq
& & A_{m,s}
-2\rho|d_{m}(\omega_{s})|\,g\,\mathbf{1}^{\mathsf{T}}\mathbf{r}_{m}(\omega_{s})
+4\rho|d_{m}(\omega_{s})|\,\mathbf{r}_{m}^{\mathsf{T}}(\omega_{s})\mathbf{u}
\nonumber\\
& &\quad
+\rho^{2}q_{0,m,s}\,t
-4\rho^{2}\mathbf{q}_{m,s}^{\mathsf{T}}\mathbf{z}
+4\rho^{2}\sum_{i,j}Q_{m,s,ij}Z_{ij}.
\end{IEEEeqnarray*}

After expressing each stochastic feasibility condition in the linear form
\eqref{eq:lin-ineq-sample}, the next objective is to assemble all these
sample-wise constraints into a single, tractable optimization program.
Because the decision variables include both binary reflection coefficients and
continuous amplifier parameters, a mixed-integer convex formulation is
required to capture their coupling.

\subsection{Mixed-Integer Conic Feasibility (at fixed $\tau$)}
To achieve this integration, we introduce the substitutions
$t=g^{2}$, $b_i=2y_i-1$, and $y_i\!\in\!\{0,1\}$. These mappings linearize the quadratic dependence on $g$ and represent the binary reflection states in a continuous form. To handle the remaining bilinear products, auxiliary variables are defined as $s_{ij}=y_i y_j$, $u_i=g y_i$, $z_i=t y_i$, and $Z_{ij}=t s_{ij}$.   Under these definitions, each $\mathcal E_s(\cdot;\tau)$ in \eqref{eq:bigM}–\eqref{eq:lin-ineq-sample} becomes *affine* in the extended variable set $(\mathbf y,\mathbf s,\mathbf u,g,t,\mathbf z,\mathbf Z)$. This reformulation converts the stochastic nonlinear problem into a form suitable for modern conic optimization solvers.

At this stage, the stochastic feasibility constraints have been fully linearized and expressed in affine form.   To render the design computationally tractable, we now proceed through three structured steps. First, we apply convex relaxations to the bilinear terms and enforce the
quadratic relations through conic constraints.   Next, we assemble all scenario-wise inequalities into a unified sample-average (SAA) feasible set that can be queried by a solver.  Finally, we describe the bisection-based MISOCP oracle that searches for the  maximum achievable SINR threshold satisfying the prescribed reliability level.

\paragraph*{Convex envelopes (McCormick) and SOC link.}
Once the bilinear terms are made explicit, convex envelopes are applied to
bound them tightly within known variable ranges.  The McCormick envelopes for $s_{ij}$, $u_i$, $z_i$, and $Z_{ij}$ remain valid because these variables are sample-independent and satisfy the uniform bounds
$y_i\!\in\![0,1]$, $g\!\in\![0,g_{\max}]$, and
$t\!\in\![0,g_{\max}^{2}]$.  
The quadratic relation $t=g^{2}$ is handled via a rotated SOC relaxation,
\begin{gather}
\|(2g,\ t-1)\|_2\le t+1,\qquad
0\le g\le g_{\max},\quad 0\le t\le g_{\max}^{2},
\label{eq:soc-stoch}
\end{gather}
which guarantee $t\ge g^{2}$ while preserving convexity and numerical stability. Accordingly, the resulting MISOCP solves a convexified outer approximation of the original bilinear model; the final $(\mathbf b,g)$ must be verified through out-of-sample Monte Carlo testing and can be tightened by re-solving with $t$ fixed to $g^2$.

\paragraph*{SAA chance-feasible set.}
With all convex relaxations in place, the entire stochastic design can be written as a single feasibility set that collects all scenario-wise constraints, binary indicators, and convex-linking constraints. For a fixed SINR target \(\tau\), define
\[
\mathbf{x}\triangleq(\mathbf{y},g,t,\mathbf{s},\mathbf{u},\mathbf{z},\mathbf{Z},\mathbf{v}).
\]

\begin{equation}
\label{eq:Cs}
\mathcal{C}_{S}(\tau)\triangleq
\left\{
\mathbf{x}\ \middle|\
\begin{array}{l}
\mathcal{E}_{s}(\mathbf{x};\tau)\ge-\mathsf{M}_{\mathrm{big}}v_{s},\quad s=1,\ldots,S,\\[2pt]
\sum_{s=1}^{S} v_{s}\le \kappa,\\[2pt]
\text{all McCormick envelope constraints hold},\\
\text{the rotated-SOC linking constraint holds},\\[2pt]
\mathbf{y}\in\{0,1\}^{N},\ \mathbf{v}\in\{0,1\}^{S}
\end{array}
\right\}.
\end{equation}

The rotated-SOC linking constraint is given in \eqref{eq:soc-stoch}.

\paragraph*{Bisection on $\tau$ and MISOCP oracle.}
Because the feasibility of $\mathcal C_S(\tau)$ is monotonic in $\tau$ (i.e., if $\mathcal C_S(\tau)$ is feasible, then $\mathcal C_S(\tau')$ is feasible for any $\tau'\le\tau$), the largest reliability-guaranteed threshold can be obtained efficiently via scalar bisection. Algorithm~\ref{alg:saa_misocp} summarizes this procedure: at each iteration, we set $\tau$ to the midpoint of $[\tau_{\mathrm L},\tau_{\mathrm U}]$ and solve the SAA--MISOCP feasibility problem over $\mathcal C_S(\tau)$. Each feasibility check enforces the sample-wise SINR constraints for all but at most $\kappa=\lfloor\varepsilon S\rfloor$ scenarios, thereby implementing the outage budget of the original chance constraint. If the problem is feasible, $\tau_{\mathrm L}$ is increased, and the incumbent solution $(\mathbf y,g)$ is stored; otherwise, $\tau_{\mathrm U}$ is decreased.
In practice, warm-starting $\tau_{\mathrm U}$ using a passive (e.g., $g=0$) or heuristic RIS configuration substantially accelerates convergence and reduces solver runtime.

\begin{algorithm}[t]
\caption{SAA-based Reliability Max--SINR via Bisection and MISOCP Feasibility Oracle}
\label{alg:saa_misocp}
\DontPrintSemicolon
\KwIn{Samples $\{\omega_s\}_{s=1}^S$; outage level $\varepsilon$ with $\kappa=\lfloor\varepsilon S\rfloor$; gain cap $g_{\max}$ from~\eqref{eq:gmax-stochastic}; bisection tolerance $\varepsilon_{\tau}$; warm-start upper bound $\tau_{\mathrm U}^{(0)}$ (e.g., passive/heuristic).}
\KwOut{$\tau^\star,\ \mathbf b^\star,\ g^\star$}

Initialize $\tau_{\mathrm L}\gets 0$, $\tau_{\mathrm U}\gets \tau_{\mathrm U}^{(0)}$\;
Initialize $(\mathbf y^{\mathrm{best}},g^{\mathrm{best}})\gets (\mathbf 0,0)$\;

\While{$\tau_{\mathrm U}-\tau_{\mathrm L}>\varepsilon_{\tau}$}{
  $\tau \gets (\tau_{\mathrm L}+\tau_{\mathrm U})/2$\;

  \tcp{MISOCP feasibility check (oracle): find variables in $\mathcal C_S(\tau)$}
  Solve the feasibility problem: find $(\mathbf y,g,t,\mathbf s,\mathbf u,\mathbf z,\mathbf Z,\mathbf v)\in\mathcal C_S(\tau)$\;

  \eIf{feasible}{
      $\tau_{\mathrm L}\gets\tau$\;
      Store $(\mathbf y^{\mathrm{best}},g^{\mathrm{best}})\gets(\mathbf y,g)$\;
  }{
      $\tau_{\mathrm U}\gets\tau$\;
  }
}

$\tau^\star \gets \tau_{\mathrm L}$\;
$\mathbf b^\star \gets 2\,\mathbf y^{\mathrm{best}}-\mathbf{1}$\;
$g^\star \gets g^{\mathrm{best}}$\;

\tcp*{Optional: refine $g$ for fixed $\mathbf b^\star$ by 1-D search on $[0,g_{\max}]$.}
\end{algorithm}

This procedure returns the largest $\tau$ that is feasible for the SAA-based MISOCP approximation at the prescribed violation budget $\kappa$. The obtained design is then certified through an out-of-sample Monte Carlo evaluation.

\paragraph*{Remarks.}
In practice, a conservative yet safe value of $\mathsf{M}_{\mathrm{big}}$ can be obtained by bounding
both sides of \eqref{eq:feas-int-stoch} at $g_{\max}$ using the eigenvalue limits of $\mathbf Q(\omega_s)$ and the norms of $\mathbf r(\omega_s)$, which balances numerical robustness and relaxation tightness; overly large $\mathsf{M}_{\mathrm{big}}$ values can weaken the MILP/MISOCP relaxation and harm solver scalability. If computational complexity becomes critical, the binary indicators in the chance constraint may be replaced by a convex Conditional Value-at-Risk (CVaR) surrogate, which avoids integer variables at the cost of a softer probabilistic guarantee. Finally, McCormick products $(s_{ij},Z_{ij})$ need only to be instantiated for indices where $Q_{ij}\neq 0$ occurs in any scenario, significantly reducing the solver’s dimensionality.
 
\paragraph*{Complexity considerations.}
Relative to the deterministic oracle, the SAA-based MISOCP introduces $S$ additional affine constraints of the form \eqref{eq:lin-ineq-sample} (one per scenario) and $S$ binary indicator variables $\{v_s\}_{s=1}^{S}$, together with the violation-budget constraint $\sum_{s=1}^{S}v_s\le\kappa$ where $\kappa=\lfloor \varepsilon S\rfloor$. The McCormick envelopes and the rotated SOC link \eqref{eq:soc-stoch} are sample-independent and are therefore shared across all scenarios. Consequently, the scenario layer scales linearly with $S$, whereas the McCormick linearization scales on the order of $N^{2}$ due to the pairwise products required by $\mathbf b^{\mathsf T}\mathbf Q(\omega_s)\mathbf b$ (and similarly for the interferers). In practice, warm-starting $\tau$ from a passive RIS design (e.g., $g=0$) or from an alternating-optimization baseline significantly accelerates convergence and yields stable runtimes across Monte-Carlo realizations.

\paragraph*{Example instance size.}
For the largest setting used in Sec.~\ref{sec:5} (i.e., $N=128$, $S=200$, and $\varepsilon=0.1$, hence $\kappa=20$), each SAA--MISOCP feasibility instance contains $N+S=328$ binary variables (the RIS states $\mathbf y\in\{0,1\}^{N}$ and the scenario indicators $\mathbf v\in\{0,1\}^{S}$). Assuming a dense $\mathbf Q(\omega_s)$ so that all bilinear terms are instantiated, the reformulation introduces approximately $2N^{2}+2N+2=33{,}026$ continuous variables (including $\mathbf s,\mathbf Z,\mathbf u,\mathbf z,g,t$) and about $8N^{2}+8N+S+1=132{,}297$ linear constraints, plus one rotated second-order cone enforcing $t\ge g^{2}$ and simple bound constraints. If the symmetry/sparsity of $\mathbf Q(\omega_s)$ is exploited (e.g., keeping only the upper triangle and omitting zero entries), the $O(N^{2})$ counts reduce proportionally. Finally, the overall runtime is proportional to the number of bisection iterations, which is $I_{\tau}\approx\left\lceil\log_{2}\!\big((\tau_{\mathrm U}-\tau_{\mathrm L})/\varepsilon_{\tau}\big)\right\rceil$ feasibility solves.

\section{\label{sec:4}Performance Bounds and Interference-Aware Analysis}

This section derives \emph{realization-wise} deterministic envelopes that bound the instantaneous
$\mathrm{SINR}(\mathbf b,g)$ for all channel realizations and establishes an interference-aware
high-gain SINR ceiling. These closed-form limits translate the stochastic model of
Sec.~\ref{sec:2} into interpretable constraints, revealing how the active gain $g$, the RIS size $N$,
and the cochannel-interference structure fundamentally cap the achievable SINR.

Building on these bounds, we characterize the achievable performance limits and robustness of the proposed system under stochastic channels and gain-dependent amplifier noise. All large- and small-scale propagation effects, including $\mathbf H_r(\omega)$ and $\boldsymbol\Sigma_a(g,\omega)$, are modeled as random variables.

The instantaneous SINR for any binary RIS reflection pattern $\mathbf b$ and amplifier gain $g\in[0,g_{\max}]$ is given in \eqref{eq:sinr-compact-stoch}. This ratio captures the combined contributions of direct, reflected, and interfering paths, along with the stochastic effects of receiver and amplifier noise.

To obtain tractable performance limits, the coupling coefficients are bounded using norm and eigenvalue inequalities:
\begin{align}
|B(\mathbf b,\omega)|&\le 2\rho|d(\omega)|\,\|\mathbf r(\omega)\|_1,\nonumber\\
\rho^2\lambda_{\min}\!\big(\mathbf Q(\omega)\big)\,N
&\le C(\mathbf b,\omega)\le\rho^2\lambda_{\max}\!\big(\mathbf Q(\omega)\big)\,N.
\label{eq:BC-bounds}
\end{align}
Analogously, for each interferer $m$,
\begin{align}
|B_m(\mathbf b,\omega)|&\le 2\rho|d_m(\omega)|\,\|\mathbf r_m(\omega)\|_1,\nonumber\\
\rho^2\lambda_{\min}\!\big(\mathbf Q_m(\omega)\big)\,N
&\le C_m(\mathbf b,\omega)\le\rho^2\lambda_{\max}\!\big(\mathbf Q_m(\omega)\big)\,N.
\label{eq:BCm-bounds}
\end{align}

For compactness, define
\begin{align}
\bar B(\omega)&\triangleq 2\rho|d(\omega)|\,\|\mathbf r(\omega)\|_1,\nonumber\\
\underline C(\omega)&\triangleq \rho^2N\lambda_{\min}\!\big(\mathbf Q(\omega)\big),\quad
\overline C(\omega)\triangleq \rho^2N\lambda_{\max}\!\big(\mathbf Q(\omega)\big),\nonumber\\
\bar B_m(\omega)&\triangleq 2\rho|d_m(\omega)|\,\|\mathbf r_m(\omega)\|_1,\nonumber\\
\underline C_m(\omega)&\triangleq \rho^2N\lambda_{\min}\!\big(\mathbf Q_m(\omega)\big),\quad
\overline C_m(\omega)\triangleq \rho^2N\lambda_{\max}\!\big(\mathbf Q_m(\omega)\big).
\label{eq:envelope-defs}
\end{align}
and
\begin{align}
\underline N(g,\omega)&\triangleq A(\omega)-g\,\bar B(\omega)+g^2\underline C(\omega),\nonumber\\
\overline N(g,\omega)&\triangleq A(\omega)+g\,\bar B(\omega)+g^2\overline C(\omega),
\label{eq:N-defs}
\end{align}
\begin{align}
\underline D(g,\omega)&\triangleq D_0(\omega)+g^2D_1(\omega)\nonumber\\
&\quad+\sum_{m=1}^{M}P_m\!\left(A_m(\omega)-g\,\bar B_m(\omega)+g^2\underline C_m(\omega)\right),\nonumber\\
\overline D(g,\omega)&\triangleq D_0(\omega)+g^2D_1(\omega)\nonumber\\
&\quad+\sum_{m=1}^{M}P_m\!\left(A_m(\omega)+g\,\bar B_m(\omega)+g^2\overline C_m(\omega)\right).
\label{eq:D-defs}
\end{align}

Substituting these bounds into \eqref{eq:sinr-compact-stoch} yields deterministic realization-wise envelopes:
\begin{align}
\underline{\mathrm{SINR}}(g;\omega)
&=
\frac{P_{\mathrm d}\,\underline N(g,\omega)}{\overline D(g,\omega)},
\label{eq:LB-stoch}
\\
\overline{\mathrm{SINR}}(g;\omega)
&=
\frac{P_{\mathrm d}\,\overline N(g,\omega)}{\underline D(g,\omega)}.
\label{eq:UB-stoch}
\end{align}

Hence, over the admissible gain interval where the lower-envelope denominator is positive,
\[
\underline{\mathrm{SINR}}(g;\omega)\le \mathrm{SINR}(\mathbf b,g;\omega)\le \overline{\mathrm{SINR}}(g;\omega),
\quad \forall\,\mathbf b\in\{\pm1\}^N.
\]
This condition is naturally satisfied in physically meaningful operating points because thermal noise contributes a strictly positive term in \(D_0(\omega)\). A full derivation is provided in Appendix~B.

When \(g\) is small, numerical stability can be improved by applying a
triangle-inequality bound:
\[
\big(\sqrt{A}-\sqrt{C_{\max}}\,g\big)_{+}^{2}
\le A+gB+g^2C
\le\big(\sqrt{A}+\sqrt{C_{\max}}\,g\big)^{2},
\]
where \(C_{\max}=\rho^{2}N\lambda_{\max}(\mathbf Q)\), with analogous expressions for interferers.


\noindent\textbf{High-gain SINR ceiling (interference-aware saturation).}
A key implication of the gain-dependent folded-noise model is that \emph{increasing $g$
cannot improve SINR beyond a finite limit}. Specifically, as $g$ grows, the quadratic
reflection energies $g^{2}C(\mathbf b,\omega)$ and $\{g^{2}C_m(\mathbf b,\omega)\}$ dominate the
numerator and the reflected-interference terms, while the folded amplifier-noise contribution
$g^{2}D_1(\omega)$ simultaneously dominates the denominator. Dividing the SINR expression by $g^{2}$
therefore yields the realization-wise ceiling
\begin{gather}
\lim_{g\to\infty}\mathrm{SINR}(\mathbf{b},g;\omega)=\frac{P_{\mathrm d}\,C(\mathbf{b},\omega)}{D_{1}(\omega)+\sum_{m}P_{m}\,C_{m}(\mathbf{b},\omega)}\nonumber \\
\ \le\ \frac{P_{\mathrm d}\,\rho^{2}\lambda_{\max}(\mathbf{Q}(\omega))N}{D_{1}(\omega)+\sum_{m}P_{m}\,\rho^{2}\lambda_{\min}(\mathbf{Q}_{m}(\omega))N}.\label{eq:ceiling-stoch}
\end{gather} Notably, the direct-path and cross terms $A(\omega)$ and $B(\mathbf b,\omega)$ vanish in this limit; thus, the high-gain behavior is governed solely by $\{C,C_m\}$ and $D_1(\omega)$. 

Active amplification is therefore beneficial compared with passive reflection whenever the asymptotic ceiling in \eqref{eq:ceiling-stoch} exceeds the passive SINR
\[
\frac{P_{\mathrm d}A(\omega)}{D_0(\omega)+\sum_{m=1}^{M}P_mA_m(\omega)}.
\]
A sufficient condition for this improvement is
\begin{gather}
\frac{P_{\mathrm d}\rho^{2}\lambda_{\max}\!\big(\mathbf Q(\omega)\big)N}
{D_{1}(\omega)+\sum_{m=1}^{M}P_m\,\rho^{2}\lambda_{\min}\!\big(\mathbf Q_m(\omega)\big)N}
>
\frac{P_{\mathrm d}A(\omega)}
{D_{0}(\omega)+\sum_{m=1}^{M}P_mA_m(\omega)}.
\label{eq:beneficial-stoch}
\end{gather}

Averaging or percentile evaluation of this inequality across channel realizations provides design guidelines for the expected performance benefits or outages that result from the operation of the active RIS.

The pathwise envelopes \eqref{eq:LB-stoch}–\eqref{eq:UB-stoch} and the high-gain ceiling \eqref{eq:ceiling-stoch} describe how the instantaneous SINR varies across random realizations of the RIS channel and amplifier noise. However, these scenario-specific results do not, by themselves, guarantee a prescribed level of reliability when the environment fluctuates. To translate these findings into a robust and certifiable design framework, the analysis proceeds in three stages: (i) we formulate a reliability-aware optimization that enforces the SINR constraint with a specified probability; (ii) we extract interpretable scaling laws that reveal how physical parameters govern stochastic performance; and (iii) we outline verification metrics to certify and interpret the obtained design.
\subsubsection*{(i) Chance-Constrained Design (Recap)}
The complete reliability-aware optimization framework is presented in Sec.~\ref{sec:3}. 
There, we maximize the guaranteed SINR threshold $\tau$ under the chance constraint
\[
\Pr\{\mathrm{SINR}(\mathbf b,g;\omega)\ge\tau\}\ge 1-\varepsilon,
\]
with $\mathbf b\in\{\pm1\}^N$ and $0\le g\le g_{\max}$. The probabilistic constraint is approximated by SAA using $S$ scenarios and violation budget $\kappa=\lfloor\varepsilon S\rfloor$, and solved via bisection with an MISOCP feasibility oracle (Algorithm~\ref{alg:saa_misocp}).
Accordingly, this section does not repeat those derivations; instead, it uses the resulting solution $(\mathbf b^\star,g^\star,\tau^\star)$ to interpret the envelope bounds in \eqref{eq:LB-stoch}--\eqref{eq:UB-stoch} and the high-gain ceiling in \eqref{eq:ceiling-stoch}.

\subsubsection*{(ii) Interpretable Scaling Laws (Stochastic Regime)}
Before moving to numerical validation, it is useful to extract analytical trends
from the stochastic envelopes.  
Equations \eqref{eq:LB-stoch}–\eqref{eq:UB-stoch} and
\eqref{eq:ceiling-stoch} reveal how system parameters influence reliable SINR.
\begin{itemize}
  \item \textbf{RIS size ($N$):}
  Desired and interfering reflections scale respectively with
  $\rho^2 N\lambda_{\max}(\mathbf Q)$ and
  $\rho^2 N\lambda_{\min}(\mathbf Q_m)$.
  Increasing $N$ widens the SINR range but can magnify interference if
  eigenstructures align.
  \item \textbf{Eigenstructure:}
  Strong alignment of the desired LoS component
  (large $\lambda_{\max}(\mathbf Q)$)
  and weak alignment of interferers
  (small $\lambda_{\min}(\mathbf Q_m)$)
  elevate the SINR ceiling and enhance robustness.
  \item \textbf{Gain-dependent noise:}
  The excess-noise coefficient $D_1(\omega)$ limits the SINR ceiling at large $g$.
  Strong folding gain $L(\omega)$ or correlated amplifier noise
  (large eigenvalues of $\boldsymbol\Sigma_{\mathrm{ex}}(\omega)$)
 reduces the benefit of active gain.
  \item \textbf{Outage sensitivity:}
  Stricter reliability (smaller $\varepsilon$) shifts the feasible $\tau$
  toward the lower envelope \eqref{eq:LB-stoch}, illustrating the trade-off between
  guaranteed SINR and allowable risk.
\end{itemize}

\subsubsection*{(iii) Verification and Reporting Metrics}
To assess the robustness and interpretability of the obtained design,
Monte–Carlo testing is performed on an independent test set
$\{\tilde{\omega}_r\}_{r=1}^{S_{\mathrm{test}}}$ using the optimized parameters
$(\mathbf b^\star,g^\star,\tau^\star)$.  
Three complementary diagnostic metrics are employed:

\begin{enumerate}
\item \textbf{Empirical reliability:}
The achieved reliability is estimated as
\[
\widehat{p}_{\mathrm{succ}}
=\frac{1}{S_{\mathrm{test}}}\sum_{r=1}^{S_{\mathrm{test}}}
  \mathbb 1\{\mathrm{SINR}(\mathbf b^\star,g^\star;\tilde{\omega}_r)\ge\tau^\star\}.
\]
which approximates the true non-outage probability.
\item \textbf{SINR statistics:}
Report the empirical mean, variance, and $(1-\varepsilon)$-quantile of
$\mathrm{SINR}(\mathbf b^\star,g^\star;\omega_s)$, confirming that $\tau^\star$
lies near the target quantile boundary.
\item \textbf{Ceiling gap:}
Evaluate the scenario-wise ceiling \eqref{eq:ceiling-stoch} to quantify how far the achieved SINR operates below the theoretical high-gain limit.
\end{enumerate}
Together, these diagnostics validate the probabilistic design guarantees and link the stochastic optimization outcomes to the physical RIS–satellite model developed in Sections~\ref{sec:2}–\ref{sec:3}.

\section{\label{sec:5}Numerical Results and Discussions}

This section validates the stochastic framework by confronting its analytical envelopes, gain ceilings, and chance-constrained design with Monte--Carlo evidence generated under the Rician block-fading channel model in \eqref{eq:rician-stoch}, the instantaneous SINR model in \eqref{eq:sinr-compact-stoch}, and the folded-noise law in \eqref{eq:D0D1-stoch}. We first list the parameters used consistently across the figures; then, we build the discussion progressively: we begin with a quantile performance view versus the interference load (validating \textbf{C1}), then move to the gain–reliability tradeoff in a per-panel analysis that reveals the ceiling mechanism (validating \textbf{C3}), and finally quantify the reliability-guaranteed surface over $(N,M)$ at fixed gains (validating \textbf{C2}, \textbf{C4}). Throughout, the admissible gain interval strictly respects the stability–EIRP cap \eqref{eq:gmax-stochastic} instantiated via \eqref{eq:gEIRP-wc}–\eqref{eq:gEIRP-cantelli}.

Unless otherwise stated, all simulations follow the stochastic model in Sec.~\ref{sec:2} with Rician block fading \eqref{eq:rician-stoch}, folded-noise coefficients
\eqref{eq:D0D1-stoch}, and the admissible gain cap $g_{\max}$ from \eqref{eq:gmax-stochastic}.
We use $S=200$ i.i.d. stochastic realizations per configuration and target outage
$\varepsilon=0.1$ (i.e., $90\%$ non-outage), so the SAA violation budget is
$\kappa=\lfloor \varepsilon S\rfloor=20$.
Table~\ref{tab:params-results} lists all default parameters and sweep ranges.

To connect the normalized parameters to a representative satellite link budget, we
consider a Ka-band downlink with a carrier frequency $f_c=20$~GHz and a bandwidth
$B=100$~MHz, as well as a LEO altitude (satellite–ground distance scale) of $h=400$~km.
The receiver thermal-noise power over bandwidth $B$ is $N_{\mathrm{th}}=kT B$,
and with a noise figure $F$ (linear, corresponding to $\mathrm{NF}$ in dB), the effective
noise becomes $N_{\mathrm{eff}}=kTB\,F$.
Throughout the simulations, we normalize the receiver noise variance to $N_0=1$;
thus, all reported SINR values can be mapped to absolute units by scaling transmit
powers by the factor $1/N_{\mathrm{eff}}$ for the chosen $(B,\mathrm{NF})$.
This normalization keeps the stochastic SINR structure \eqref{eq:sinr-compact-stoch}
unchanged while enabling the physical interpretation of the results.

\begin{table}[!t]
\centering
\footnotesize
\renewcommand{\arraystretch}{1.15}
\setlength{\tabcolsep}{5pt}
\caption{Simulation parameters used in Sec.~\ref{sec:5} (defaults and sweeps unless stated).}
\label{tab:params-results}
\begin{tabularx}{\columnwidth}{@{}l l X@{}}
\toprule
\textbf{Symbol} & \textbf{Parameter} & \textbf{Value / Sweep} \\
\midrule
$f_c$ & Carrier frequency & $20$~GHz (Ka-band example) \\
$B$ & System bandwidth & $100$~MHz \\
$h$ & LEO altitude / distance scale & $400$~km \\
$\mathrm{NF}$ & Receiver noise figure & $5$~dB (used for physical interpretation; simulations use $N_0=1$ normalization) \\
\midrule
$N$ & RIS elements & $\{16,32,64,128\}$ \\
$M$ & Cochannel interferers & $\{2,4,6,8\}$ \\
$g$ & Active gain (common) & Panels $\{0,0.5,1,2\}$; sweeps $[0,\,g_{\max}]$ \\
$\rho$ & Passive retention factor & $0.9$ \\
$P_{\mathrm d}$ & Desired-link power & $P_{\mathrm d}=1$ \\
$P_m$ & Interferer power & $P_m=P_{\mathrm d}$ \\
$N_0$ & Receiver noise variance & $1$ (normalized) \\
$\sigma_{\min}^2$ & RIS amplifier baseline noise & $0.05$ \\
$\eta$ & Excess-noise coefficient & $0.02$ \\
$K$ & Rician $K$-factor & $6$ (linear) \\
$S$ & SAA sample size & $200$ \\
$\varepsilon$ & Outage probability & $0.10$ \\
\bottomrule
\end{tabularx}
\end{table}

\begin{figure}[t]
\begin{centering}
\includegraphics[width=3.5in,viewport=2bp 0bp 550bp 350bp]{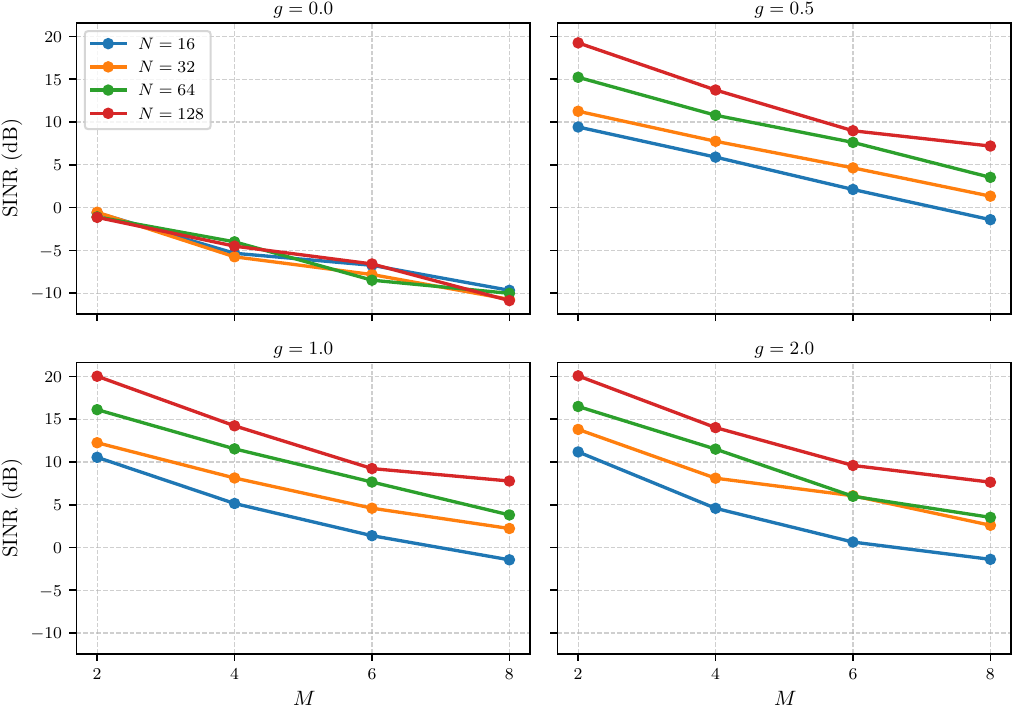}
\par\end{centering}
\caption{$(1-\varepsilon)$-quantile SINR versus number of interferers $M$ at fixed gains $g$ (panels) and RIS sizes $N$ (curves), computed from \eqref{eq:sinr-compact-stoch} under the SAA sampling settings in Table~\ref{tab:params-results}. The passive baseline ($g=0$) follows directly from \eqref{eq:sinr-compact-stoch}, while saturation at larger $g$ is consistent with the high-gain ceiling in \eqref{eq:ceiling-stoch}.}

\label{fig:quantile-vs-M}
\end{figure}

We begin with Fig.~\ref{fig:quantile-vs-M}, which plots the empirical $(1\!-\!\varepsilon)$–quantile of $\mathrm{SINR}$ as a function of the interference load $M$ for several fixed gains $g$ (panels) and RIS sizes $N$ (curves). When $g=0$, \eqref{eq:sinr-compact-stoch} reduces to the passive ratio $\mathrm{SINR}=P_{\mathrm d}A/\!\big(D_{0}+\sum_{m}P_{m}A_{m}\big)$ with $D_{0}$ given by \eqref{eq:D0D1-stoch}, which is essentially independent of $N$; the only trend is the monotonic degradation with $M$ because the denominator increases additively. This directly reflects the lower/upper envelopes in \eqref{eq:LB-stoch}–\eqref{eq:UB-stoch} specialized to $g=0$ and thus validates the \textbf{C1} claim that the eigenvalue/$\ell_1$ bounds control the entire SINR range path-wise. As soon as $g>0$, the reflected terms $g\,B(\mathbf b,\omega)$ and $g^{2}C(\mathbf b,\omega)$ enter the numerator, while $g^{2}D_1(\omega)$ appears in the denominator. Because $C(\mathbf b,\omega)=\rho^2\mathbf b^{\mathsf T}\mathbf Q(\omega)\mathbf b$ is confined by $\rho^2 N\lambda_{\min}(\mathbf Q)\le C\le \rho^2 N\lambda_{\max}(\mathbf Q)$ from \eqref{eq:BC-bounds}, enlarging $N$ reliably shifts the curves upward. Yet, the improvement saturates as $g$ grows because \eqref{eq:ceiling-stoch} limits the high-gain behavior by the ratio of $C$-terms to $D_{1}+\sum_m P_m C_m$, specifically the flattening observed in the active panels. The figure therefore combines the two pillars of \textbf{C1}: (i) path-wise envelopes \eqref{eq:LB-stoch}–\eqref{eq:UB-stoch} tracking the entire gain region, and (ii) the asymptotic ceiling \eqref{eq:ceiling-stoch} explaining saturation under gain-dependent noise.

\begin{figure}[t]
\begin{centering}
\includegraphics[width=3.5in,viewport=2bp 0bp 550bp 350bp]{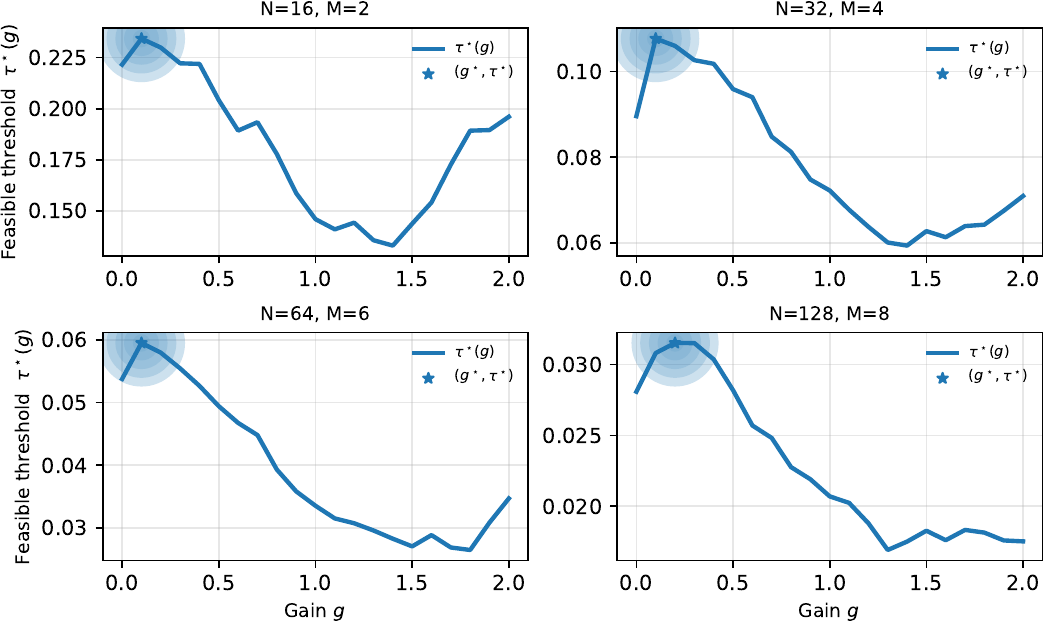}
\par\end{centering}
\caption{Reliability-guaranteed SINR threshold $\tau^\star(g)$ obtained by the SAA-MISOCP bisection oracle (Algorithm~\ref{alg:saa_misocp}) for \eqref{eq:chance-prob}, with sample-wise constraints \eqref{eq:feas-int-stoch} enforced via \eqref{eq:bigM}. The saturation/decline trend at large $g$ is explained by the folded-noise term $g^2D_1(\omega)$ in \eqref{eq:D0D1-stoch} and the ceiling behavior in \eqref{eq:ceiling-stoch}.}
\label{fig:tau-vs-g}
\end{figure}

The saturation observed at higher gains in Fig.~\ref{fig:quantile-vs-M} motivates a reliability-centered interpretation, shown in Fig.~\ref{fig:tau-vs-g}. For each fixed $(N,M)$ panel and each gain value $g$, we solve the SAA--MISOCP feasibility problem and report $\tau^\star(g)$ as the largest SAA-feasible SINR threshold under the violation budget $\kappa=\lfloor\varepsilon S\rfloor$ in \eqref{eq:bigM}. Equivalently, on the SAA training set, at most $\kappa$ out of $S$ scenarios violate the SINR constraint, so the empirical non-outage level is at least $1-\kappa/S\approx 1-\varepsilon$ (up to solver tolerances and the Big-$M$/McCormick/SOC reformulation). Therefore, at finite $S$, \eqref{eq:chance-prob} is targeted empirically rather than certified exactly; the final reliability is validated using independent out-of-sample Monte-Carlo testing.

The trend in $\tau^\star(g)$ follows directly from \eqref{eq:sinr-compact-stoch}. For small $g$, the numerator is mainly $A(\omega)+gB(\mathbf b,\omega)$ while the denominator changes slowly, so choosing $\mathbf b$ aligned with $\mathbf r(\omega)$ makes $B(\mathbf b,\omega)$ positive and yields an initial near-linear increase. As $g$ grows, both desired and interference reflection terms scale as $g^2$ through $C(\mathbf b,\omega)$ and ${C_m(\mathbf b,\omega)}$, and the folded-noise term $g^2D_1(\omega)$ also increases in the denominator (see \eqref{eq:D0D1-stoch}). Consequently, $\tau^\star(g)$ eventually saturates and may decline, consistent with the ceiling law in \eqref{eq:ceiling-stoch}. Practically, this identifies a useful operating-gain region: beyond it, additional amplification mainly increases internal noise rather than reliable SINR. This behavior supports \textbf{C3} and remains consistent with the stochastic-envelope interpretation in \textbf{C1}.

\begin{figure}[t]
\begin{centering}
\includegraphics[width=3.8in,viewport=2bp 0bp 550bp 350bp]{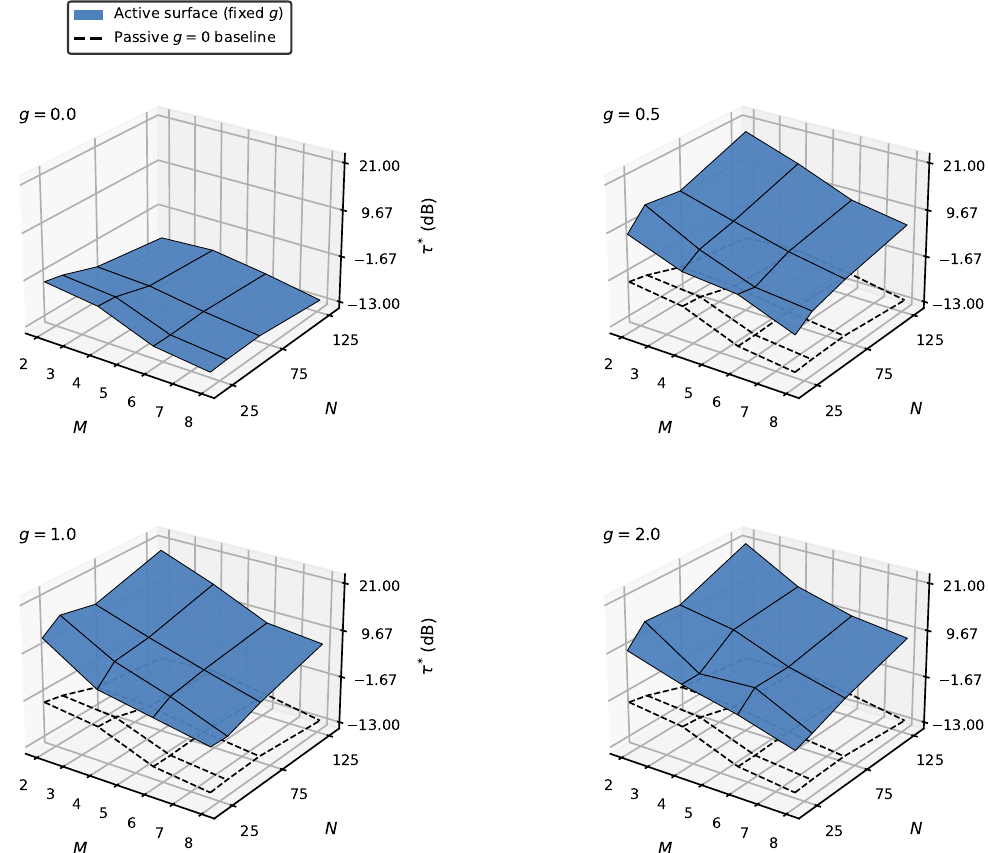}
\par\end{centering}
\caption{Reliability-guaranteed SINR surface $\tau^\star(N,M;g)$ at fixed gains $g\in\{0,0.5,1,2\}$ obtained for \eqref{eq:chance-prob} using the SAA--MISOCP model with sample-wise constraints \eqref{eq:feas-int-stoch} enforced through \eqref{eq:bigM}. The scaling with $N$ follows the quadratic-reflection term $C(\mathbf b,\omega)$ bounded in \eqref{eq:BC-bounds}, while degradation with $M$ follows the additive interference structure in \eqref{eq:sinr-compact-stoch}.}

\label{fig:tau-surface}
\end{figure}

Having seen how reliability evolves with $g$ for a given $(N,M)$, Fig.~\ref{fig:tau-surface} broadens the view of the entire $(N,M)$ plane at fixed gains. Each surface point is the SAA $(1-\varepsilon)$-feasible threshold $\tau^\star(N,M;g)$ obtained from \eqref{eq:chance-prob} using the sample-wise constraints \eqref{eq:feas-int-stoch} (linearized as \eqref{eq:lin-ineq-sample}) and enforced through \eqref{eq:bigM}.

Increasing $N$ lifts the surface due to coherent scaling $C(\mathbf b,\omega)\sim \rho^2 N\lambda_{\max}(\mathbf Q(\omega))$ in \eqref{eq:BC-bounds}, whereas increasing $M$ depresses it through the additive interference term $\sum_{m=1}^{M}P_m\!\left(A_m+gB_m+g^2C_m\right)$ in \eqref{eq:sinr-compact-stoch}. The dashed wireframe at $g=0$ provides the passive baseline. The clear separation between the active ($g>0$) and passive ($g=0$) surfaces visualizes how the proposed stochastic design enables \emph{controlled amplification}: active gain improves the reliability-guaranteed SINR only within the physically admissible range $0\le g\le g_{\max}$ imposed by stability and EIRP constraints in \eqref{eq:gmax-stochastic}. Regions where the active surfaces lie above the passive wireframe satisfy the sufficient improvement condition \eqref{eq:beneficial-stoch}, confirming that amplification is beneficial precisely when the high-gain ceiling in \eqref{eq:ceiling-stoch} exceeds the passive ratio $P_{\mathrm d}A/D_0$. Since each $(N,M;g)$ point is obtained from the SAA violation budget (with $\kappa=\lfloor\varepsilon S\rfloor$), the figure illustrates the reliability-targeting design and its out-of-sample validation, as well as the \textbf{C4} physics--optimization bridge: the SAA constraints \eqref{eq:feas-int-stoch} are enforced via the Big-$M$ violation-budget formulation \eqref{eq:bigM} with McCormick/SOC convexification, while respecting the gain cap \eqref{eq:gmax-stochastic}.

\begin{figure}[t]
\begin{centering}
\includegraphics[width=3.5in,viewport=2bp 0bp 550bp 350bp]{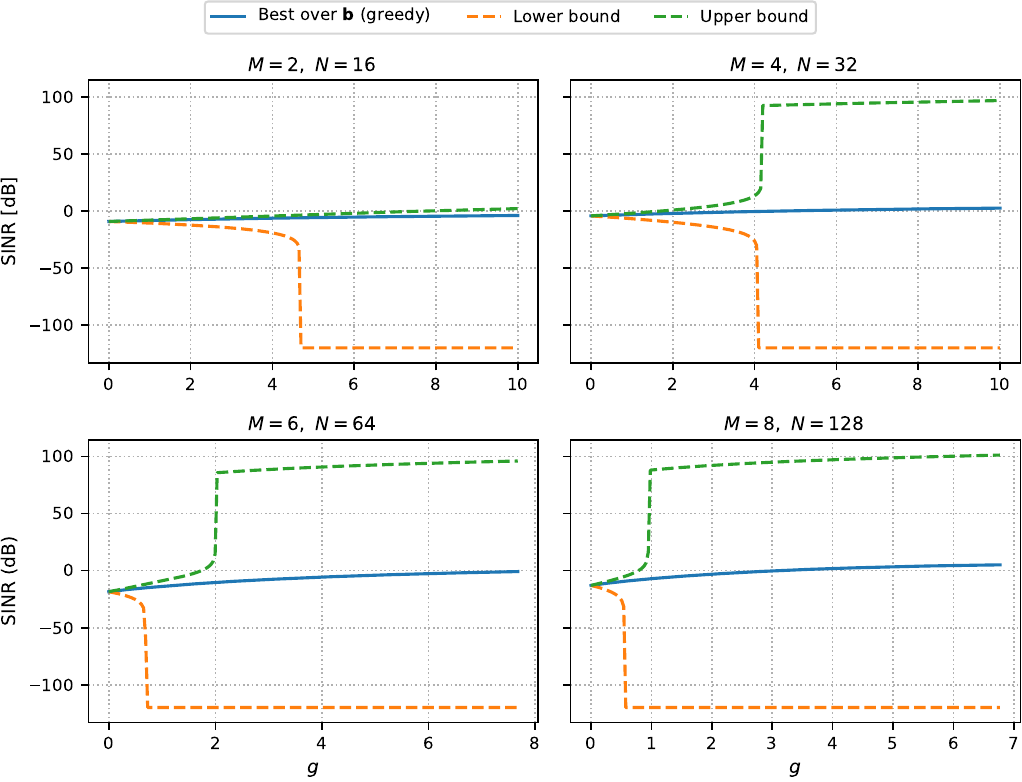}
\par\end{centering}
\caption{Envelope consistency across gains: for each realization $\omega$, the path-wise bounds \eqref{eq:LB-stoch}--\eqref{eq:UB-stoch} provably sandwich $\mathrm{SINR}(\mathbf b,g;\omega)$ for all $\mathbf b$ and $g\in[0,g_{\max}]$. The plotted empirical “best over $\mathbf b$” curve therefore lies between the two envelopes, validating the bound tightness and the ceiling trend \eqref{eq:ceiling-stoch}.}

\label{fig:envelopes}
\end{figure}

Finally, Fig.~\ref{fig:envelopes} closes the loop by checking that the empirical “best over $\mathbf b$” curve is indeed sandwiched between the deterministic envelopes predicted by \eqref{eq:LB-stoch}–\eqref{eq:UB-stoch} for all gains in the admissible range $0\le g\le g_{\max}$. At small $g$, triangle-inequality regularization (as stated after \eqref{eq:UB-stoch}) guarantees nonnegativity and explains the gentle initial slope, while at larger $g$, all three curves flatten in unison toward the ceiling \eqref{eq:ceiling-stoch}. Because the horizontal axis truncation reflects $g_{\max}$ from \eqref{eq:gmax-stochastic} (with $g_{\mathrm{EIRP}}$ selected via \eqref{eq:gEIRP-wc}, \eqref{eq:gEIRP-quantile}, or \eqref{eq:gEIRP-cantelli} depending on the reliability posture), the comparison remains physically meaningful and numerically stable. This figure, therefore, provides a direct, per-realization validation of \textbf{C1} while reinforcing the ceiling phenomenon of \textbf{C3} and the feasibility architecture of \textbf{C4}.

Together, Figs.~\ref{fig:quantile-vs-M}--\ref{fig:envelopes} validate the model: the SINR envelopes \eqref{eq:LB-stoch}--\eqref{eq:UB-stoch} and ceiling \eqref{eq:ceiling-stoch} explain the trends over $(N,M,g)$, while folded noise \eqref{eq:D0D1-stoch} explains saturation at high $g$. The reliability gains predicted by \eqref{eq:beneficial-stoch} are observed, and the chance-constrained SAA--MISOCP design in \eqref{eq:chance-prob} (implemented via \eqref{eq:feas-int-stoch}, \eqref{eq:lin-ineq-sample}, and \eqref{eq:bigM}) matches Monte-Carlo results across passive/active regimes and low/high interference.

\section{\label{sec:6}Conclusion}
This paper proposes a stochastic reliability framework for active RIS--assisted satellite downlinks under random fading, cochannel interference, and gain-dependent (folded) amplifier noise. Reliability is enforced through a chance constraint on the instantaneous SINR and is solved via an SAA-based MISOCP feasibility oracle combined with scalar bisection, while explicitly restricting the common gain to the physically admissible interval $0\le g\le g_{\max}$ determined by stability and EIRP limits.

Analytically, we derived realization-wise SINR envelopes and an interference-aware high-gain ceiling that explain the transition from useful amplification to saturation: increasing $g$ strengthens the reflected component only until the quadratic folded-noise term $g^{2}D_{1}(\omega)$ (together with interference-related $g^{2}C_m(\mathbf b,\omega)$ terms) dominates the denominator. This yields a quantifiable \emph{reliability-gain trade-off} and effectively characterizes the usable gain region for active RIS operation under realistic noise statistics.

Monte-Carlo results corroborated the theory by showing that the envelopes track the simulated SINR behavior across passive and active regimes, capture the observed saturation at higher gains, and reproduce the expected degradation with increasing interference. The resulting reliability-guaranteed surfaces further highlight the separation between passive ($g=0$) and active ($g>0$) operation, demonstrating controlled amplification without violating the stability/EIRP-imposed gain cap.

In terms of scalability, each bisection iteration requires one SAA–MISOCP feasibility solve; the scenario layer scales linearly with the sample size $S$, while the McCormick linearization scales on the order of $N^{2}$ due to pairwise bilinear products. Warm-starting from a passive or alternating-optimization baseline reduces iterations and stabilizes runtime across Monte-Carlo realizations. Future work will extend the framework to wideband and multiuser settings, finer phase/gain control, adaptive data-driven operation under time-varying CSI, and experimental validation with hardware prototypes and measured channels.

\appendices

\section{Derivation of the Quadratic-Form Coefficients}
\label{app:sinr-derivation}

This appendix provides a complete derivation of the intermediate steps that lead from the
received-signal model \eqref{eq:y-stoch} to the quadratic-form coefficient definitions
\eqref{eq:def-ABC-stoch}--\eqref{eq:def-ABCm-stoch} and to the folded-noise structure used in
\eqref{eq:sinr-compact-stoch}. The aim is to justify the identities
\begin{align}
|d(\omega)+\rho g\,\mathbf u^{\mathsf T}(\omega)\mathbf b|^2
&=A(\omega)+g\,B(\mathbf b,\omega)+g^2\,C(\mathbf b,\omega), \label{eq:app:goal1}\\
|d_m(\omega)+\rho g\,\mathbf u_m^{\mathsf T}(\omega)\mathbf b|^2
&=A_m(\omega)+g\,B_m(\mathbf b,\omega)+g^2\,C_m(\mathbf b,\omega), \label{eq:app:goal2}
\end{align}
as well as the form of the effective noise variance at the combiner output.

\subsection{Scalar equivalent model at the combiner output}
Starting from \eqref{eq:y-stoch}, define the effective desired and interfering scalar channels
\begin{align}
h(\omega) &\triangleq d(\omega)+\rho g\,\mathbf u^{\mathsf T}(\omega)\mathbf b,\label{eq:app:h}\\
h_m(\omega) &\triangleq d_m(\omega)+\rho g\,\mathbf u_m^{\mathsf T}(\omega)\mathbf b,\qquad m=1,\ldots,M,\label{eq:app:hm}
\end{align}
and the effective noise term
\begin{align}
n_{\mathrm{eff}}(\omega)\triangleq 
\mathbf w^{\mathsf H}\mathbf n_r
+\mathbf w^{\mathsf H}\mathbf H_r(\omega)\mathbf n_{\mathrm{RA}}(\omega).\label{eq:app:neff}
\end{align}
Then \eqref{eq:y-stoch} is equivalently written as
\begin{align}
y(\omega)=h(\omega)x+\sum_{m=1}^{M}h_m(\omega)x_m+n_{\mathrm{eff}}(\omega).\label{eq:app:y}
\end{align}

\subsection{Instantaneous desired/interference powers}
Because $x\sim\mathcal{CN}(0,P_{\mathrm d})$ and $x_m\sim\mathcal{CN}(0,P_m)$ are mutually independent,
and independent of the noise processes, the conditional second moments satisfy
\begin{align}
\mathbb E\!\left[|h(\omega)x|^2\mid\omega\right] &= P_{\mathrm d}|h(\omega)|^2,\label{eq:app:Px}\\
\mathbb E\!\left[|h_m(\omega)x_m|^2\mid\omega\right] &= P_m|h_m(\omega)|^2,\qquad m=1,\ldots,M.\label{eq:app:Pxm}
\end{align}
Thus, it remains to express $|h(\omega)|^2$ and $|h_m(\omega)|^2$ in a convenient quadratic form.

\subsection{Folded-noise variance at the combiner output}
The receiver noise satisfies $\mathbf n_r\sim\mathcal{CN}(\mathbf 0,N_0\mathbf I)$, hence
\begin{align}
\mathrm{Var}\!\left(\mathbf w^{\mathsf H}\mathbf n_r\right)=N_0\|\mathbf w\|_2^2.\label{eq:app:vn_r}
\end{align}
For the RIS amplifier noise, \eqref{eq:fold-general-stoch} gives
\begin{align}
\mathrm{Var}\!\left(\mathbf w^{\mathsf H}\mathbf H_r(\omega)\mathbf n_{\mathrm{RA}}(\omega)\right)
=\mathbf w^{\mathsf H}\mathbf H_r(\omega)\boldsymbol\Sigma_a(g,\omega)\mathbf H_r^{\mathsf H}(\omega)\mathbf w.\label{eq:app:vn_ra}
\end{align}
Assuming $\mathbf n_r$ and $\mathbf n_{\mathrm{RA}}(\omega)$ are independent, the total effective noise variance is
\begin{align}
\mathrm{Var}\!\left(n_{\mathrm{eff}}(\omega)\right)
&=\mathrm{Var}\!\left(\mathbf w^{\mathsf H}\mathbf n_r\right)
+\mathrm{Var}\!\left(\mathbf w^{\mathsf H}\mathbf H_r(\omega)\mathbf n_{\mathrm{RA}}(\omega)\right)\nonumber\\
&=N_0\|\mathbf w\|_2^2
+\mathbf w^{\mathsf H}\mathbf H_r(\omega)\boldsymbol\Sigma_a(g,\omega)\mathbf H_r^{\mathsf H}(\omega)\mathbf w.\label{eq:app:vn_total}
\end{align}

\paragraph*{i.i.d.\ special case.}
When $\boldsymbol\Sigma_a(g,\omega)=\sigma_a^2(g)\mathbf I$ with $\sigma_a^2(g)=\sigma_{\min}^2+\eta g^2$ and
$L(\omega)=\|\mathbf H_r^{\mathsf H}(\omega)\mathbf w\|_2^2$, we have
\begin{align}
\mathbf w^{\mathsf H}\mathbf H_r(\omega)\boldsymbol\Sigma_a(g,\omega)\mathbf H_r^{\mathsf H}(\omega)\mathbf w
&=\sigma_a^2(g)\,\mathbf w^{\mathsf H}\mathbf H_r(\omega)\mathbf H_r^{\mathsf H}(\omega)\mathbf w\nonumber\\
&=\left(\sigma_{\min}^2+\eta g^2\right)L(\omega).\label{eq:app:iid_fold}
\end{align}
Substituting \eqref{eq:app:iid_fold} into \eqref{eq:app:vn_total} yields
\begin{align}
\mathrm{Var}\!\left(n_{\mathrm{eff}}(\omega)\right)=
\underbrace{N_0\|\mathbf w\|_2^2+\sigma_{\min}^2L(\omega)}_{D_0(\omega)}
+\underbrace{\eta L(\omega)}_{D_1(\omega)}g^2,\label{eq:app:D0D1}
\end{align}
which is the $D_0(\omega)+D_1(\omega)g^2$ structure used in the denominator of \eqref{eq:sinr-compact-stoch}.

\subsection{Phase-aligned expansion of the desired effective channel power}
We now derive the quadratic representation of $|h(\omega)|^2$ in \eqref{eq:app:goal1}.
Let $\phi(\omega)=\arg(d(\omega))$ and define the phase-rotated vector
\begin{align}
\tilde{\mathbf u}(\omega)\triangleq \mathbf u(\omega)e^{-j\phi(\omega)}.\label{eq:app:utilde}
\end{align}
Since $d(\omega)e^{-j\phi(\omega)}=|d(\omega)|$, the magnitude is invariant under a common phase rotation:
\begin{align}
|h(\omega)|^2
&=\left|d(\omega)+\rho g\,\mathbf u^{\mathsf T}(\omega)\mathbf b\right|^2\nonumber\\
&=\left|e^{-j\phi(\omega)}\!\left(d(\omega)+\rho g\,\mathbf u^{\mathsf T}(\omega)\mathbf b\right)\right|^2\nonumber\\
&=\left||d(\omega)|+\rho g\,\tilde{\mathbf u}^{\mathsf T}(\omega)\mathbf b\right|^2.\label{eq:app:rotinv}
\end{align}
Expanding the squared magnitude gives
\begin{align}
|h(\omega)|^2
&=|d(\omega)|^2
+2\rho g|d(\omega)|\,\Re\!\left\{\tilde{\mathbf u}^{\mathsf T}(\omega)\mathbf b\right\}
+\rho^2 g^2\left|\tilde{\mathbf u}^{\mathsf T}(\omega)\mathbf b\right|^2.\label{eq:app:expand1}
\end{align}

Define the real-valued quantities (as in the main text)
\begin{align}
\mathbf r(\omega)&\triangleq \Re\{\tilde{\mathbf u}(\omega)\},\label{eq:app:rdef}\\
\mathbf Q(\omega)&\triangleq \Re\!\left\{\tilde{\mathbf u}(\omega)\tilde{\mathbf u}^{\mathsf H}(\omega)\right\}.
\label{eq:app:Qdef}
\end{align}
Then the cross term satisfies
\begin{align}
\Re\!\left\{\tilde{\mathbf u}^{\mathsf T}(\omega)\mathbf b\right\}
=\sum_{i=1}^{N} b_i\,\Re\!\left\{\tilde u_i(\omega)\right\}
=\mathbf b^{\mathsf T}\mathbf r(\omega).\label{eq:app:cross}
\end{align}
Moreover, since $\mathbf b\in\{\pm 1\}^N$ is real,
\begin{align}
\left|\tilde{\mathbf u}^{\mathsf T}(\omega)\mathbf b\right|^2
&=\left(\tilde{\mathbf u}^{\mathsf T}(\omega)\mathbf b\right)\left(\tilde{\mathbf u}^{\mathsf T}(\omega)\mathbf b\right)^{\!*}\nonumber\\
&=\tilde{\mathbf u}^{\mathsf T}(\omega)\mathbf b\,\mathbf b^{\mathsf T}\tilde{\mathbf u}^{*}(\omega)\nonumber\\
&=\mathbf b^{\mathsf T}\left(\tilde{\mathbf u}(\omega)\tilde{\mathbf u}^{\mathsf H}(\omega)\right)\mathbf b\nonumber\\
&=\mathbf b^{\mathsf T}\Re\!\left\{\tilde{\mathbf u}(\omega)\tilde{\mathbf u}^{\mathsf H}(\omega)\right\}\mathbf b
=\mathbf b^{\mathsf T}\mathbf Q(\omega)\mathbf b.\label{eq:app:quad}
\end{align}
Because $\mathbf b\in\mathbb R^N$, the quadratic form is real and nonnegative:
\[
\mathbf b^{\mathsf T}\mathbf Q(\omega)\mathbf b
=\left|\tilde{\mathbf u}^{\mathsf T}(\omega)\mathbf b\right|^2\ge0,
\] which is the PSD property required by the optimization model.
Substituting \eqref{eq:app:cross} and \eqref{eq:app:quad} into \eqref{eq:app:expand1} yields
\begin{align}
|h(\omega)|^2
=|d(\omega)|^2
+2\rho g|d(\omega)|\,\mathbf b^{\mathsf T}\mathbf r(\omega)
+\rho^2 g^2\,\mathbf b^{\mathsf T}\mathbf Q(\omega)\mathbf b.\label{eq:app:desiredABC}
\end{align}
Identifying
\begin{align}
A(\omega)&=|d(\omega)|^2,\nonumber\\
B(\mathbf b,\omega)&=2\rho|d(\omega)|\,\mathbf b^{\mathsf T}\mathbf r(\omega),\nonumber\\
C(\mathbf b,\omega)&=\rho^2\,\mathbf b^{\mathsf T}\mathbf Q(\omega)\mathbf b,\nonumber
\end{align}
recovers the definitions in \eqref{eq:def-ABC-stoch} and establishes \eqref{eq:app:goal1}.

\subsection{Phase-aligned expansion of each interferer power}
For interferer $m$, define $\phi_m(\omega)=\arg(d_m(\omega))$ and
\begin{align}
\tilde{\mathbf u}_m(\omega)\triangleq \mathbf u_m(\omega)e^{-j\phi_m(\omega)}.\label{eq:app:utilde_m}
\end{align}
Repeating the steps in \eqref{eq:app:rotinv}--\eqref{eq:app:quad}, define
\begin{align}
\mathbf r_m(\omega)&\triangleq \Re\{\tilde{\mathbf u}_m(\omega)\},\\
\mathbf Q_m(\omega)&\triangleq \Re\!\left\{\tilde{\mathbf u}_m(\omega)\tilde{\mathbf u}_m^{\mathsf H}(\omega)\right\}\succeq\mathbf 0,
\end{align}
and obtain
\begin{align}
|h_m(\omega)|^2
=|d_m(\omega)|^2
+2\rho g|d_m(\omega)|\,\mathbf b^{\mathsf T}\mathbf r_m(\omega)
+\rho^2 g^2\,\mathbf b^{\mathsf T}\mathbf Q_m(\omega)\mathbf b.\label{eq:app:interfABC}
\end{align}
Identifying
\begin{align}
A_m(\omega)&=|d_m(\omega)|^2,\nonumber\\
B_m(\mathbf b,\omega)&=2\rho|d_m(\omega)|\,\mathbf b^{\mathsf T}\mathbf r_m(\omega),\nonumber\\
C_m(\mathbf b,\omega)&=\rho^2\,\mathbf b^{\mathsf T}\mathbf Q_m(\omega)\mathbf b,\nonumber
\end{align}
recovers the definitions in \eqref{eq:def-ABCm-stoch} and establishes \eqref{eq:app:goal2}.

Equations \eqref{eq:app:goal1}-\eqref{eq:app:goal2} yield the coefficient forms used in
\eqref{eq:def-ABC-stoch}-\eqref{eq:def-ABCm-stoch}, and \eqref{eq:app:D0D1} provides the folded-noise
structure used in \eqref{eq:sinr-compact-stoch}.

 \section{Derivation of Realization-Wise SINR Envelopes}
\label{app:sinr-envelopes}

This appendix derives the deterministic realization-wise lower and upper SINR envelopes used in \eqref{eq:LB-stoch}--\eqref{eq:UB-stoch}.

For a fixed realization \(\omega\), RIS state \(\mathbf b\in\{\pm1\}^N\), and gain \(g\ge 0\), the instantaneous SINR in \eqref{eq:sinr-compact-stoch} can be written as
\begin{align}
\mathrm{SINR}(\mathbf b,g;\omega)
=
\frac{P_{\mathrm d}\,\mathcal N(\mathbf b,g;\omega)}
{\mathcal D(\mathbf b,g;\omega)},
\label{eq:appB-sinr}
\end{align}
where
\begin{align}
\mathcal N(\mathbf b,g;\omega)
&\triangleq A(\omega)+gB(\mathbf b,\omega)+g^2C(\mathbf b,\omega),
\label{eq:appB-num}
\\
\mathcal D(\mathbf b,g;\omega)
&\triangleq D_0(\omega)+g^2D_1(\omega)\nonumber\\
&\quad+\sum_{m=1}^{M}P_m\!\left(A_m(\omega)+gB_m(\mathbf b,\omega)+g^2C_m(\mathbf b,\omega)\right).
\label{eq:appB-den}
\end{align}

Using \eqref{eq:BC-bounds}--\eqref{eq:BCm-bounds} and the definitions in \eqref{eq:envelope-defs}, we have
\begin{align}
-\bar B(\omega)\le B(\mathbf b,\omega)\le \bar B(\omega),\qquad
\underline C(\omega)\le C(\mathbf b,\omega)\le \overline C(\omega),
\label{eq:appB-BC-des}
\end{align}
and, for each interferer \(m\),
\begin{align}
-\bar B_m(\omega)\le B_m(\mathbf b,\omega)\le \bar B_m(\omega),\qquad
\underline C_m(\omega)\le C_m(\mathbf b,\omega)\le \overline C_m(\omega).
\label{eq:appB-BC-int}
\end{align}

Hence, for the desired term,
\begin{align}
A(\omega)-g\bar B(\omega)+g^2\underline C(\omega)
\le \mathcal N(\mathbf b,g;\omega)
\le A(\omega)+g\bar B(\omega)+g^2\overline C(\omega).
\label{eq:appB-num-bounds}
\end{align}
Using \eqref{eq:N-defs}, this is
\begin{align}
\underline N(g,\omega)\le \mathcal N(\mathbf b,g;\omega)\le \overline N(g,\omega).
\label{eq:appB-num-bounds-compact}
\end{align}

Similarly, for the denominator terms,
\begin{align}
A_m(\omega)-g\bar B_m(\omega)+g^2\underline C_m(\omega)
&\le A_m(\omega)+gB_m(\mathbf b,\omega)+g^2C_m(\mathbf b,\omega)\nonumber\\
&\le A_m(\omega)+g\bar B_m(\omega)+g^2\overline C_m(\omega).
\label{eq:appB-int-term-bounds}
\end{align}
Multiplying by \(P_m\ge 0\), summing over \(m\), and adding \(D_0(\omega)+g^2D_1(\omega)\), we obtain
\begin{align}
\underline D(g,\omega)\le \mathcal D(\mathbf b,g;\omega)\le \overline D(g,\omega),
\label{eq:appB-den-bounds}
\end{align}
where \(\underline D(g,\omega)\) and \(\overline D(g,\omega)\) are defined in \eqref{eq:D-defs}.

Assume the admissible operating region satisfies \(\underline D(g,\omega)>0\). Then all relevant denominators are positive, and ratio monotonicity yields
\begin{align}
\frac{\underline N(g,\omega)}{\overline D(g,\omega)}
\le
\frac{\mathcal N(\mathbf b,g;\omega)}{\mathcal D(\mathbf b,g;\omega)}
\le
\frac{\overline N(g,\omega)}{\underline D(g,\omega)}.
\label{eq:appB-ratio}
\end{align}
Multiplying by \(P_{\mathrm d}\ge 0\) and using \eqref{eq:LB-stoch}--\eqref{eq:UB-stoch}, we obtain
\begin{align}
\underline{\mathrm{SINR}}(g;\omega)
\le
\mathrm{SINR}(\mathbf b,g;\omega)
\le
\overline{\mathrm{SINR}}(g;\omega),
\qquad \forall\,\mathbf b\in\{\pm1\}^N.
\label{eq:appB-final}
\end{align}
This completes the derivation.
\hfill\(\blacksquare\)

\bibliographystyle{IEEEtran}
\bibliography{SNR}

\end{document}